\documentclass[useAMS,referee,usenatbib]{biom}

\usepackage{setspace}
\def\bSig\mathbf{\Sigma}

\usepackage{amsmath}
\usepackage{amsfonts}
\usepackage{graphicx}
\usepackage{xcolor}
\usepackage{mathtools}

\title[Dynamic factor model for ILD]{A Continuous-Time Dynamic Factor Model for Intensive Longitudinal Data Arising from Mobile Health Studies}

\author
{Madeline R. Abbott\emailx{mrabbott@umich.edu} \\
Department of Biostatistics, University of Michigan--Ann Arbor, Ann Arbor, Michigan, U.S.A.
\and
Walter H. Dempsey\emailx{wdem@umich.edu} \\
Department of Biostatistics, University of Michigan--Ann Arbor, Ann Arbor, Michigan, U.S.A.
\and
Inbal Nahum-Shani\emailx{inbal@umich.edu} \\
Institute for Social Research, University of Michigan--Ann Arbor, Ann Arbor, Michigan, U.S.A.
\and
Cho Y. Lam\emailx{cho.lam@hci.utah.edu} \\
Department of Population Health Sciences and Huntsman Cancer Institute, University of Utah, \\ Salt Lake City, UT, U.S.A.
\and
David W. Wetter\emailx{david.wetter@hci.utah.edu} \\
Department of Population Health Sciences and Huntsman Cancer Institute, University of Utah, \\ Salt Lake City, UT, U.S.A.
\and
Jeremy M. G. Taylor\emailx{jmgt@umich.edu} \\
Department of Biostatistics, University of Michigan--Ann Arbor, Ann Arbor, Michigan, U.S.A.}

\begin{document}

\date{   }

\pagerange{\pageref{firstpage}--\pageref{lastpage}} 
  
\label{firstpage}

\begin{abstract}
Intensive longitudinal data (ILD) collected in mobile health (mHealth) studies contain rich information on multiple outcomes measured frequently over time that have the potential to capture short-term and long-term dynamics. Motivated by an mHealth study of smoking cessation in which participants self-report the intensity of many emotions multiple times per day, we describe a dynamic factor model that summarizes the ILD as a low-dimensional, interpretable latent process. This model consists of two submodels: (i) a measurement submodel---a factor model---that summarizes the multivariate longitudinal outcome as lower-dimensional latent variables and (ii) a structural submodel---an Ornstein-Uhlenbeck (OU) stochastic process---that captures the temporal dynamics of the multivariate latent process in continuous time. We derive a closed-form likelihood for the marginal distribution of the outcome and the computationally-simpler sparse precision matrix for the OU process. We propose a block coordinate descent algorithm for estimation. Finally, we apply our method to the mHealth data to summarize the dynamics of 18 different emotions as two latent processes. These latent processes are interpreted by behavioral scientists as the psychological constructs of positive and negative affect and are key in understanding vulnerability to lapsing back to tobacco use among smokers attempting to quit.
\end{abstract}


\maketitle

\section{Introduction}
\label{s:intro}

Intensive longitudinal data (ILD) can capture rapid changes in outcomes over time. In mobile health (mHealth) studies, information about multiple longitudinal outcomes is often collected with the aim of understanding the temporal dynamics of unobservable constructs related to mental or physical health. Our work is motivated by an mHealth study of smoking cessation in which the intensity of emotions over time was collected from current smokers attempting to quit. Participants self-reported the intensity of 18 different emotions up to four times per day over 10 days, resulting in a substantial quantity of rich data. For smoking cessation researchers, understanding the temporal dynamics of the latent psychological states that underlie these emotions is of scientific interest.

The volume and complexity of ILD, however, make them challenging to analyze since longitudinal outcomes are often measured irregularly across many individuals; thus statistical methods must be able to handle the high volume of irregularly spaced data. At the same time, the frequent measurements in ILD create many opportunities to discover new information, particularly if the latent constructs of interest vary rapidly. We present a dynamic factor model that is motivated by the need to model multiple longitudinal outcomes measured frequently over time in a flexible yet interpretable manner. Our proposed model consists of two submodels: (i) a measurement submodel---a factor model---that summarizes the multiple observed longitudinal outcomes as lower-dimensional latent factors and (ii) a structural submodel---an Ornstein-Uhlenbeck (OU) stochastic process---that captures the evolution of the multiple correlated latent factors over time. Together, these components of our dynamic factor model are flexible enough to capture the variability in the longitudinal outcomes while avoiding use of a non-parametric or other many-parameter model that inhibits interpretability. In addition to improving interpretability, the low-dimensional nature of the structural submodel also greatly reduces computational complexity, as opposed to fitting a high-dimensional stochastic process directly to the observed outcomes.

One standard approach to modeling changes in multiple correlated longitudinal variables over time is to use an autoregressive (AR) model. These models, which are called vector autoregressive (VAR) models when data are multivariate, have been widely used to model observed outcomes as well as latent variables. For example, \citet*{dunson_2003}, \citet*{cui_2014}, and \citet{tran_2019} have proposed related methods in which observed longitudinal outcomes are summarized as time-varying lower-dimensional latent variables. The correlation of these latent variables is then modeled with AR or VAR processes. VAR models, however, are specified for balanced data. This situation is often not realistic in the case of ILD, which generally consists of irregularly-measured outcomes, and can lead to biased estimates in cases where the assumption is made but does not hold.

Mixed models have been proposed as alternatives to discrete-time processes for modeling the evolution of latent variables over time and have been previously used in combination with factor models. Unlike the AR and VAR processes, mixed models do not require balanced data. Existing work has focused both on the development of mixed models for modeling the evolution of a single latent factor over time (e.g., \citealp{roy_2000}; \citealp{proust_2006}; \citealp*{proust_lima_2013}) or multiple latent factors (e.g., \citealp{liu_2019}; \citealp*{wang_2013}). Overall, these mixed model-based approaches are useful tools for capturing smooth trends in latent factors. In our application, however, we aim to develop a method that can capture the correlation between and rapid variation in multiple latent emotional constructs over time.

The OU process, which can be thought of as a continuous-time analog of the AR or VAR process, is a stochastic process well-suited for capturing rapid variations over time. Existing work has frequently focused on using the OU process or integrated OU process to model longitudinal outcomes that have been directly observed (or observed with measurement error); e.g., \citet*{taylor_1994}; \citet*{sy_1997}; \citet*{oravecz_2009}; \citet*{oravecz_2016}.

In more recent work, the OU process has also been used in the context of latent variable models. \citet{tran_2021a} propose a latent linear mixed model that summarizes multiple observed longitudinal outcomes as a smaller number of latent factors while accounting for the serial correlation between repeated measurements over time via an OU process. This work differs from ours, however, in that the fixed effects that capture the association between the observed covariates and the latent factors are of primary interest; the OU process is incorporated into the structural mixed model as a tool for accounting for serial correlation between repeated observations.

Most closely related to our proposed approach is work by \citet{tran_2021b}. Like us, they propose a longitudinal latent variable model that consists of two parts: a measurement submodel to summarize observed outcomes as lower dimensional latent factors and an OU process as the structural submodel for the latent factors. While we differ in the exact specification of the measurement submodel, our chosen models are related. Key distinctions between this existing work and the approach presented in this manuscript are in the model parameterization and computational approach. Tran et al. (2021b) take a Bayesian approach, which requires sampling values of the latent process at each measurement occasion. In the ILD setting, we need approaches that can scale to large numbers of repeated measurements. Here, we choose to work in the frequentist framework.  As a result of taking a maximum likelihood-based approach, we can directly maximize the marginal log-likelihood of the observed longitudinal outcome.  Furthermore, this framework enables us to employ various algebraic and computational strategies to make estimation faster, resulting in a method more suitable for ILD.

In this work, we fill a gap in the existing literature by proposing an Ornstein-Uhlenbeck factor (OUF) model that captures the temporal dynamics between rapidly-varying correlated latent factors observed via multiple longitudinal outcomes and an estimation algorithm with the computational efficiency to handle ILD. Our novel contributions include (i) a closed-form likelihood for the marginal distribution of the observed outcome, (ii) the derivation of the computationally-simpler sparse precision matrix for the multivariate OU process, (iii) identifiability constraints imposed via scaling constants, and (iv) a block coordinate descent algorithm for estimation and inference in a maximum likelihood framework.

The remainder of this paper is organized as such:  In Section \ref{s:motivating_data}, we describe the motivating mHealth data; in Section \ref{s:method}, we present our novel method and contributions; in Section \ref{s:sim_study}, we demonstrate the performance of our method via simulation; in Section \ref{s:data_app}, we illustrate our model via a scientifically--meaningful application to mHealth data; and in Section \ref{s:discussion}, we provide a discussion.

\section{Motivating data}\label{s:motivating_data}

\begin{figure}
    \centering
    \includegraphics[width=\linewidth]{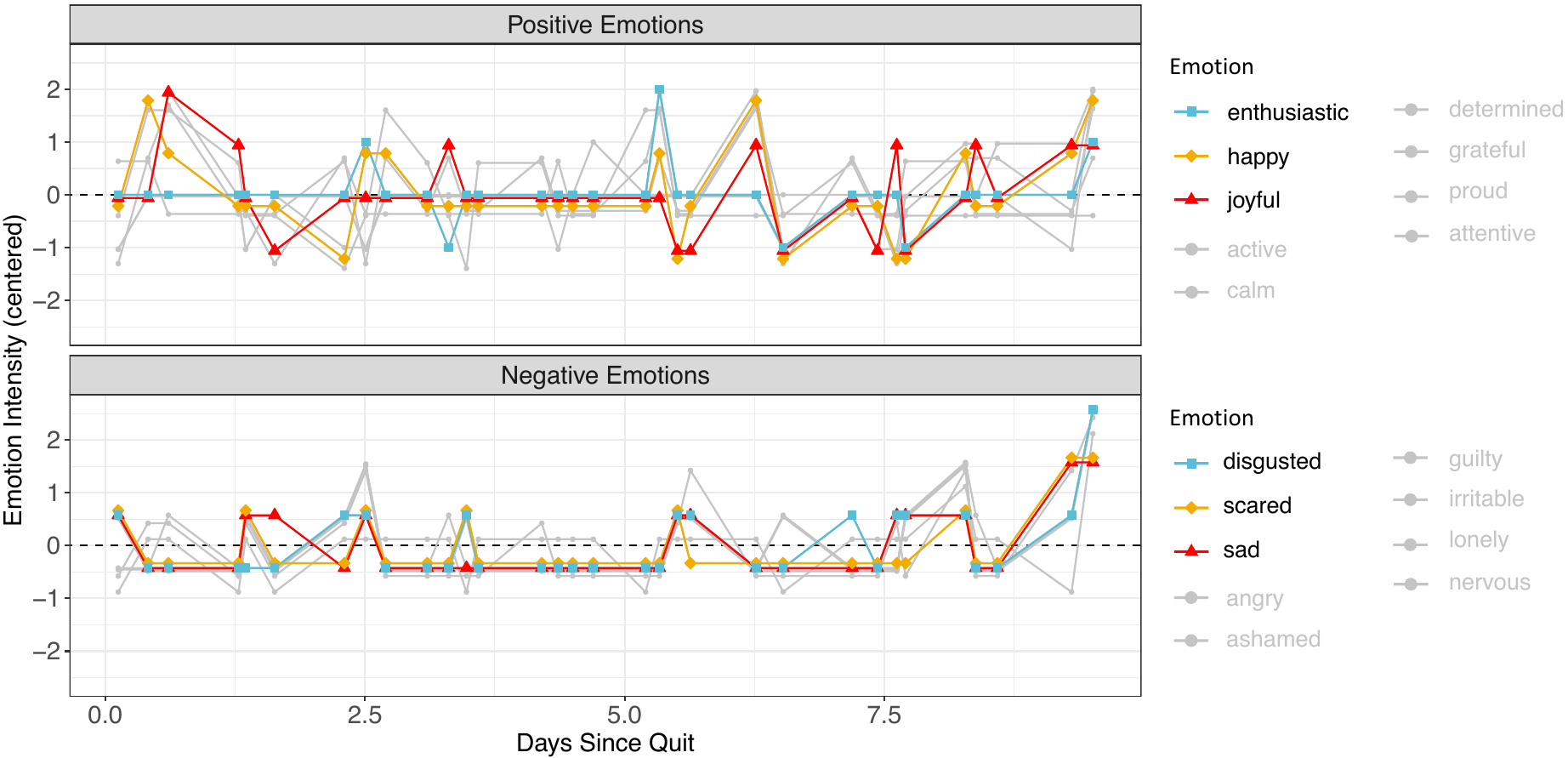}
    \caption{Responses to emotion-related questions over time for one participant in the mHealth study. In this plot, a subset of three positive emotions and three negative emotions are highlighted solely for illustrative purposes; all 18 emotions are later included in the model described in Section \ref{s:data_app}. Note the rapid fluctuations of these longitudinal outcomes over time.} \label{fig:real_dat}
\end{figure}

Data motivating this work come from an mHealth study of smoking cessation \citep{potter_2023}. In this observational study, current smokers (N = 218) who were attempting to quit were followed for 10 days. During these 10 days, ecological momentary assessments (EMAs), which enable repeated sampling of individuals' current states and contexts in real time, were used to track participants' emotions as they were experienced in a high-frequency manner. Specifically, participants were prompted to respond to a series of questions sent to their smartphones multiple per day at random occasions; the original study design intended for individuals to receive up to four EMAs per day. The EMAs contained a set of questions that assessed the current intensity of multiple emotions measured on a 5-point Likert scale. We focus on a set of 18 emotions consisting of both positive and negative emotions that attempts to capture the distinct-but-correlated underlying emotional states of positive and negative affect (i.e., summary measures of overall positive and negative feeling). The resulting data contain frequent measurements of a substantial number of longitudinal outcomes, where the number of measurement occasions per person ranges from 2 to 47 (mean = 17). Note that these data are only the subset of the full study data that were available at the time of drafting this manuscript.  Additional details on the study procedures can be found in \cite{potter_2023}. The high rate of measurement enables us to capture rapid changes in emotions over time. To illustrate the dynamics of these responses, Figure \ref{fig:real_dat} shows the responses to emotion-related EMA questions over time for one participant in the study. Understanding the dynamics of smokers' latent emotional states that underlie the measured responses is of scientific interest among smoking cessation researchers and behavioral scientists more broadly.

\section{Methods}\label{s:method}

In this section, we present the OUF model that jointly models multiple observed longitudinal outcomes and the lower dimensional latent factors assumed to generate the observed longitudinal outcomes. Our proposed model consists of two submodels: a measurement submodel and a structural submodel.

\subsection{Measurement submodel}

Let $\bm{Y}_i(t) = [Y_{i1}(t), Y_{i2}(t), ..., Y_{iK}(t)]^{\top}$ be a $K \times 1$ vector of measured longitudinal outcomes for individual $i, i = 1, ..., N$, at time $t$. Assume that individual $i$ has longitudinal outcomes measured at $n_i$ occasions. Using the measurement submodel, we model the observed longitudinal outcome $\bm{Y}_i(t)$ as
\begin{equation}
\begin{aligned}
\label{eq:meas_submodel}
    \bm{Y}_{i}(t) = \bm{\Lambda} \bm{\eta}_i(t) + \bm{u}_i + \bm{\epsilon}_i(t)
\end{aligned}
\end{equation}

\noindent where $\bm{\eta}_i(t)$ is a vector of $p$ time-varying latent factors (where $p < K$); $\bm{\Lambda}$ is a $K \times p$-dimensional time-invariant loadings matrix with elements $\lambda_{k,j}$ that captures the degree of association between the latent factors and observed longitudinal outcomes; $\bm{u}_i \sim N(0, \bm{\Sigma}_u)$ is a vector of length $K$ of random intercepts; and $\bm{\epsilon}_i(t) \sim N(0, \bm{\Sigma}_{\epsilon})$ is a vector representing measurement error, where $\bm{\Sigma}_{\epsilon}$ is assumed to be a diagonal matrix.

This model builds upon a standard factor model but also includes (i) a random intercept and (ii) a multivariate model for the evolution of the correlated latent processes $\bm{\eta}_i(t)$ (described in Section \ref{ss:structural_submod}).  We assume that $\bm{\Sigma}_u$ is diagonal, as we include this term to account for the longitudinal correlation in the repeated measurements but then model the correlated change in outcomes through the structural submodel.  Allowing a non-diagonal $\bm{\Sigma}_u$ is possible, but we opt not to do so to avoid the substantial increase computational cost associated with estimation of these extra parameters.  We also assume that $\bm{\Lambda}$ contains many structural zeros such that each row of the loadings matrix contains only one non-zero element; this structure means that each observed outcome is a measurement of only a single underlying latent factor. The decision to incorporate structural zeros in the loadings matrix is supported by behavioral science concepts (i.e., Positive and Negative Affect Schedule; PANAS \citep*{watson_1988}), which classify a given emotion as a measurement of a specific category of emotional state.  Learning the location of the structural zeros, rather than pre-specifying them, is a possible direction for future work.

\subsection{Structural submodel}\label{ss:structural_submod}

The structural submodel captures the evolution of the latent factors, $\bm{\eta}_i(t)$, over time. We use a multivariate OU process, which can be understood as a continuous-time analog of a VAR process and has the ability to capture rapid temporal variation. Here, we assume a bivariate OU process ($p = 2$) for illustrative purposes. The stochastic differential equation definition of the bivariate OU process is
\begin{equation}
\begin{aligned}
    d \begin{bmatrix} \eta_{i1}(t) \\ \eta_{i2}(t) \end{bmatrix} =  - \underbrace{\begin{bmatrix} \theta_{11} & \theta_{12} \\ \theta_{21} & \theta_{22} \end{bmatrix}}_{\coloneqq\bm{\theta}} \begin{bmatrix} \eta_{i1}(t) \\ \eta_{i2}(t) \end{bmatrix} dt + \underbrace{\begin{bmatrix} \sigma_{11} & 0 \\ 0 & \sigma_{22} \end{bmatrix}}_{\coloneqq\bm{\sigma}} d\begin{bmatrix} W_{i1}(t) \\ W_{i2}(t) \end{bmatrix}
\end{aligned}
\end{equation}

\noindent where the diagonal elements of matrix $\bm{\theta}$ capture the mean-reverting tendency of the latent factors (where the mean is assumed to be $0$) and the off-diagonal elements of $\bm{\theta}$ capture correlation between the latent factors.  The diagonal elements of $\bm{\theta}$ are required to be positive.  The matrix $\bm{\sigma}$, with elements $\sigma_{11}$ and $\sigma_{22} > 0$, describes the volatility of the process, where $W_{i1}(t)$ and $W_{i2}(t)$ are both standard Brownian motion. In general, the standard definition of the OU process allows $\bm{\sigma}$ to take non-zero values in the off-diagonal. By restricting $\bm{\sigma}$ to be a simpler diagonal matrix here, we consider the Brownian motion terms as separate noise processes for each latent variable and thus capture all correlation between the latent processes through the $\bm{\theta}$ matrix. We also require that all eigenvalues of the $\bm{\theta}$ matrix have a positive real part; this constraint ensures a mean-reverting process (see \cite{vatiwutipong_2019}).  

The multivariate stochastic process provides advantages over a simpler univariate stochastic process as it captures correlation between multiple variables over time; however, the multivariate nature does increase the complexity of the model and thus the computational burden. We address this complexity in the next section.

\subsubsection{Marginal covariance and precision matrices for the OU process}\label{ss:marginal_ou}

Rather than taking a Bayesian strategy or relying on the complete-data likelihood and taking an expectation-maximization (EM) approach to estimation, we directly maximize the likelihood of the observed data.  Direct maximization of the marginal likelihood allows us to avoid repeatedly calculating values of the latent factors at each measurement occasion (via posterior sampling in a Bayesian framework or via complex integrals in the E-step of the EM algorithm). Thus, our approach is more scalable to the ILD setting.


In order to carry out our estimation algorithm (described in Section \ref{ss:est_approach}), we require the marginal covariance matrix of the OU process. \citet{vatiwutipong_2019} present a form of the conditional variance and cross-covariance function for the OU process but provide these functions in integral forms that are not amenable to likelihood-based inference. To avoid approximations resulting from numerical integration, (i) we derive an analytic form of the conditional covariance function and (ii) we account for the additional uncertainty of an unknown initial state by deriving the analytic form of the marginal covariance function. For a stationary OU process with known initial state at time $t_0 = 0$, $\bm{\eta}(t_0)$, the conditional mean at time $t$ is $\mathbb{E}\{\bm{\eta}(t) | \bm{\eta}(t_0)\} = e^{-\bm{\theta} t} \bm{\eta}(t_0)$.  Assuming a marginal mean of 0, the conditional and marginal covariance functions follow:

\begin{lemma}\label{def:conditional_cross_cov}
The analytic form of the OU conditional covariance at times $s$ and $t$, where $s \leq t$, is
\begin{equation*} 
\begin{aligned}
\label{eq:conditional_cross_cov}
    Cov\{\eta(s), \eta(t)| \eta(t_0 = 0)\} &= vec^{-1}\Big\{(\theta \oplus \theta)^{-1} \big[ e^{s (\theta \oplus \theta)} - I \big] e^{-[t\theta \oplus s\theta]} vec\{ \sigma \sigma^{\top} \} \Big\}
\end{aligned}
\end{equation*}
\end{lemma}

\begin{lemma}\label{def:marginal_cross_cov}
The analytic form of the OU marginal covariance at times $s$ and $t$, $s \leq t$, is
\begin{equation*}
\begin{aligned}
\label{eq:marginal_cross_cov}
    Cov\{ \eta(s), \eta(t) \} = vec^{-1} \Big\{ (\theta \oplus \theta)^{-1} \big[ e^{(\theta \oplus \theta)s - (\theta t \oplus \theta s)} \big] vec\{\sigma \sigma^{\top}\} \Big\}
\end{aligned}
\end{equation*}
\end{lemma}

\noindent Here, $\oplus$ denotes the Kronecker sum, defined for square matrices $\bm{A}$ and $\bm{B}$ of sizes $a$ and $b$, respectively, as $\bm{A} \oplus \bm{B} = \bm{A} \otimes \bm{I}_b + \bm{I}_a \otimes \bm{B}$; and the $vec\{ \bm{A} \}$ operation consists of stacking the columns of matrix $\bm{A}$ into a column vector. For details on the derivations of these results, see Section A.1 and A.2 of the supplementary material.

In addition to the marginal covariance function of the OU process, we derive the precision matrix. Due to the Markov property of the OU process, the precision matrix is block tri-diagonal and thus much simpler to calculate than the dense covariance matrix.

\begin{lemma}\label{def:precision_matrix}

Let $\bm{\Omega}$ be the precision matrix of the OU process observed at $n$ occasions and define the stationary variance as $\bm{V} := vec^{-1} \big\{ (\bm{\theta} \oplus \bm{\theta})^{-1} vec\{\bm{\sigma} \bm{\sigma}^{\top}\} \big\}$. Then $\bm{\Omega}$ has the structure
\begin{equation}
\begin{aligned}
\bm{\Omega} = \begin{bmatrix}
            \bm{\Omega}_{11} & \bm{\Omega}_{12} & 0 & \cdots & 0 \\
            \bm{\Omega}_{12}^{\top} & \bm{\Omega}_{22} & \bm{\Omega}_{23} & \cdots & 0 \\
            0 & \bm{\Omega}_{23}^{\top} & \bm{\Omega}_{33} & \ddots & \vdots \\
            \vdots & \vdots & \ddots & \ddots & \bm{\Omega}_{n-1,n} \\
            0 & 0 & \cdots & \bm{\Omega}_{n-1,n}^{\top} & \bm{\Omega}_{nn} \\
         \end{bmatrix}
\end{aligned}
\end{equation}

\noindent and each block indexed by $j$ for $1 < j < n$ in the tri-diagonal matrix is
\begin{equation}
\begin{aligned}
    & \bm{\Omega}_{11} = \big[ \bm{V} - \bm{V}  e^{-\bm{\theta}^{\top} (t_2 - t_1)} \bm{V}^{-1} e^{-\bm{\theta} (t_2 - t_1)} \bm{V} \big]^{-1} \\
    & \bm{\Omega}_{j,j+1} = -\big[\bm{V} - \bm{V} e^{-\bm{\theta}^{\top} (t_{j+1} - t_j)} \bm{V}^{-1} e^{-\bm{\theta} (t_{j+1} - t_j)}  \bm{V} \big]^{-1}  \bm{V} e^{-\bm{\theta}^{\top} (t_{j+1} - t_j)} \bm{V}^{-1} \\
    & \bm{\Omega}_{jj} = \bm{V}^{-1} + \bm{V}^{-1} e^{-\bm{\theta} (t_j - t_{j-1})} \bm{V} \big[\bm{V} - \bm{V} e^{-\bm{\theta}^{\top} (t_j - t_{j-1})} \bm{V}^{-1} e^{-\bm{\theta} (t_j - t_{j-1})} \bm{V} \big]^{-1} \bm{V} e^{-\bm{\theta}^{\top} (t_j - t_{j-1})} \bm{V}^{-1} \\ & \hspace{0.9cm} + \big[\bm{V} - \bm{V} e^{-\bm{\theta}^{\top} (t_{j+1} - t_j)} \bm{V}^{-1} e^{-\bm{\theta} (t_{j+1} - t_j)} \bm{V} \big]^{-1} \bm{V} e^{-\bm{\theta}^{\top} (t_{j+1} - t_j)} \bm{V}^{-1} e^{-\bm{\theta} (t_{j+1} - t_j)} \\
    & \bm{\Omega}_{nn} = \bm{V}^{-1} + \bm{V}^{-1} e^{-\bm{\theta} (t_n - t_{n-1})} \bm{V} \big[\bm{V} - \bm{V} e^{-\bm{\theta}^{\top} (t_n - t_{n-1})} \bm{V}^{-1} e^{-\bm{\theta} (t_n - t_{n-1})} \bm{V} \big]^{-1} \bm{V} e^{-\bm{\theta}^{\top} (t_n - t_{n-1})} \bm{V}^{-1}
\end{aligned}
\end{equation}
\end{lemma}

\noindent The derivation for each block is given in Section A.3 of the supplementary material. Later, during estimation, we take advantage of the sparse precision matrix to simplify computation. This sparsity becomes particularly advantageous as the number of individuals and observations per individual in a dataset increase. 

\subsection{Joint distribution and likelihood}

Together, the measurement and structural submodels imply that the observed longitudinal outcomes are normally distributed with mean 0 and covariance $\bm{\Sigma}_i^* := Var(\bm{Y}_i) = (\bm{I}_{n_i} \otimes \bm{\Lambda}) Var(\bm{\eta}_i) (\bm{I}_{n_i} \otimes \bm{\Lambda})^{\top} + \bm{J}_{n_i} \otimes \bm{\Sigma}_u + \bm{I}_{n_i} \otimes \bm{\Sigma}_{\epsilon}$, where $\bm{I}_{n_i}$ is an $n_i \times n_i$ identity matrix and $\bm{J}_{n_i}$ is an $n_i \times n_i$ matrix of ones.  We estimate the OUF model by minimizing the following function, which equal to twice the negative log-likelihood up to a constant: $-2logL(\bm{Y}) = \sum_{i = 1}^{N} log|\bm{\Sigma}_i^{*}| + \sum_{i = 1}^{N} \bm{Y}_i^{\top} \bm{\Sigma}_i^{*-1} \bm{Y}_i$.

\subsection{Identification issues}\label{ss:identification}

Before fitting our model, we must make additional assumptions to address identifiability issues common to factor models. Because both $\bm{\Lambda}$ and $\bm{\eta}_i(t)$ are unknown, multiplying $\bm{\Lambda}$ by some matrix, say $\bm{A}$, and multiplying $\bm{\eta}_i(t)$ by $\bm{A}^{-1}$ will result in the same model. To make a factor model identifiable, constraints must be placed on either the loadings matrix or the latent factors. \citet{aguilar_2000} and \citet{carvalho_2008}, for example, make the standard assumption of requiring the loadings matrix to be triangular while \citet{tran_2019}, for example, fix the variance of the latent factors at 1. We also fix the scale of the latent factors but propose a novel approach for doing so. Letting $\bm{\eta}_i$ be the $(p \times n_i)$-length vector of latent variables values stacked over measurement occasions, we constrain $Var(\bm{\eta}_i)$ to have diagonal elements equal to 1. This constraint means that the OU process must have a stationary variance equal to 1. By fixing the scale of the latent factors, we can allow the elements of the loadings matrix $\bm{\Lambda}$ to vary almost freely during estimation. For a generic $\bm{\Lambda}$ (without structural zeros), the only constraint on the loadings matrix is that the sign of the first element must be positive.  Together these constraints make our model identifiable; the constraint on the OU process identifies the scale and the constraint on the first element of the loadings matrix identifies the direction.  Because we later make the simplifying assumption that $\bm{\Lambda}$ contains structural zeros with a single non-zero loading per row, flipping the signs on both the loadings and the latent factors results in the same model; we choose to keep the signs that correspond to the most relevant interpretation of the model given the application.  Another constraint could be added to require that one loading per column of $\bm{\Lambda}$ is positive; this would avoid sign flipping.  We revisit our approach to selecting the correct sign later in the context of rescaling the OU process.

To impose this identifiability constraint, we use a set of $p$ constants to re-scale the OU process parameters. We summarize this identifiability constraint for the bivariate OU process as:
\begin{lemma}\label{def:rescale_theta_sigma}
Using a pair of positive scalar constants $c_1$ and $c_2$, we can re-scale an arbitrary OU process parameterized by $\bm{\theta}$ and $\bm{\sigma}$ to have stationary variance of 1, where this re-scaled OU process is parameterized by $\bm{\theta}^*$ and $\bm{\sigma}^*$ according to
\begin{equation}\label{eq:rescale_theta_sigma}
    \begin{bmatrix} \theta^*_{11} & \theta^*_{12} \\ \theta^*_{21} & \theta^*_{22} \end{bmatrix} = \begin{bmatrix} \theta_{11} & \frac{c_1}{c_2} \theta_{12} \\ \frac{c_2}{c_1} \theta_{21} & \theta_{22} \end{bmatrix} \text{  and  }
    \begin{bmatrix} \sigma^*_{11} & 0 \\ 0 & \sigma^*_{22} \end{bmatrix} = \begin{bmatrix} c_1 \sigma_{11} & 0 \\ 0 & c_2 \sigma_{22} \end{bmatrix}
\end{equation}
\end{lemma}

\noindent In Section A.4 of the supplementary material, we show why this re-scaling approach works for any mean-reverting OU process. This constraint can also be extended to OU processes of higher dimensions.

Although this identifiability assumption allows us to identify the magnitude of the loadings in the factor model, it does so only up to a sign change. Consider again the case of a bivariate OU process.  The likelihood for our model is equivalent for pairs of scaling constants $(c_1 = 1, c_2 = 1)$ and $(c_1 = 1, c_2 = -1)$. In practice, the model would be the same under both pairs of scaling constants (and so we restrict $c_1$ and $c_2$ to be positive during estimation) but interpretation of model parameters would differ. After estimation, the signs on estimated model parameters can easily be flipped to match the most relevant interpretation of the data by multiplying estimates of $\bm{\Lambda}$ and $\bm{\theta}$ by a $p \times p$ matrix with the constants along the diagonal.

\subsection{Estimation algorithm}\label{ss:est_approach}

To fit this model, we take an iterative approach to estimation in which we directly maximize the marginal likelihood of our observed longitudinal outcome using a block coordinate descent algorithm and rely on simpler existing models to inform the initial parameter estimates. In the block coordinate descent algorithm, we split parameters into two different blocks: one block for parameters in the measurement submodel ($\bm{\Lambda}$, $\bm{\Sigma}_{u}$, $\bm{\Sigma}_{\epsilon}$) and the other for parameters in the structural submodel ($\bm{\theta}$, $\bm{\sigma}$). Note that each element of these blocks is actually a matrix of parameters.  Within each block-wise iteration, we minimize the log-likelihood with respect to one block of parameters, given the current estimates of the other block of parameters, using Newton algorithms as implemented in \texttt{R}'s \texttt{stats} package \citep{R_stats}. By updating parameters in blocks, we can leverage the availability of analytic gradients for parameters in the measurement submodel. The Kronecker structure of the covariance matrix for each individual’s longitudinal outcomes $\bm{Y}_i$ allows us to derive these analytic gradients. We present the general structure of the gradient here:
\begin{lemma}\label{def:fa_gradients}
The gradient of the log-likelihood for a single individual with respect to one of the measurement submodel parameters, $\Theta_j$, has the general form 
\begin{equation*}
\begin{aligned}
    \frac{\partial logL(\bm{Y}_i)}{\partial\Theta_j} = -\frac{1}{2} \Bigg[ tr \Bigg\{ \Big( I - \bm{\Sigma}_i^{*-1} \bm{Y}_i \bm{Y}_i^{\top} \Big) \Sigma_i^{*-1} \frac{\partial \bm{\Sigma}_i^{*}}{\partial \Theta_j}  \Bigg\} \Bigg]
\end{aligned}
\end{equation*}
where the exact form of $\frac{\partial \bm{\Sigma}_i^{*}}{\partial \Theta_j}$ depends on the specific parameter; either $\lambda_k$, $\sigma_{u_k}$, or $\sigma_{\epsilon_k}$.
\end{lemma}

\noindent The complete set of analytic gradients is given in Section A.5 of the supplementary material.  The computational advantage of using the analytic gradient, as opposed to a numerical approach to differentiation, is particularly notable as the number of longitudinal outcomes---and thus parameters in the measurement submodel---increases.

Prior to maximizing the marginal likelihood, we use a cross-sectional factor model to initialize $\bm{\Lambda}$, $\bm{\theta}$, and $\bm{\sigma}$, and use linear mixed models to initialize $\bm{\Sigma}_u$ and $\bm{\Sigma}_{\epsilon}$. Then, we iteratively update parameter estimates using the following block coordinate descent algorithm:
\begin{enumerate}
    \item \textit{Initialize estimates of $\bm{\Lambda}^{(0)}, \bm{\Sigma}_u^{(0)}, \bm{\Sigma}_{\epsilon}^{(0)}, \bm{\theta}^{(0)}, \bm{\sigma}^{(0)}$. Measurement submodel parameters are always initialized empirically; for structural submodel parameters, two sets of initial estimates are considered---an empirical set of values estimated from cross-sectional factor scores and a default set of values.  The set of values that corresponds to the higher log-likelihood given the current data is used.}
    \item \textit{Set iteration index $r = 1$ and convergence indicator $\delta = 0$. While $\delta = 0$,}
    \begin{enumerate}
        \item \textit{Update block 1 (measurement submodel): $$\bm{\Lambda}^{(r)}, \bm{\Sigma}_u^{(r)}, \bm{\Sigma}_{\epsilon}^{(r)} = \underset{\bm{\Lambda}, \bm{\Sigma}_u, \bm{\Sigma}_{\epsilon}}{argmax}\big\{ logL(\bm{\Lambda}, \bm{\Sigma}_u, \bm{\Sigma}_{\epsilon} | Y; \bm{\theta}^{(r-1)}, \bm{\sigma}^{(r-1))}) \big\}.$$  Maximization is done via a Newton-type algorithm using analytic gradients (Lemma \ref{def:fa_gradients}).}
        \item \textit{Update block 2 (structural submodel): $$\bm{\theta}^{(r)}, \bm{\sigma}^{(r)} = \underset{\bm{\theta}, \bm{\sigma}}{argmax}\big\{ logL(\bm{\theta}, \bm{\sigma} | Y; \bm{\Lambda}^{(r)}, \bm{\Sigma}_u^{(r)}, \bm{\Sigma}_{\epsilon}^{(r)}) \big\}.$$ Maximization is done via a quasi-Newton algorithm using numerical gradients.}
        \item \textit{Using Lemma \ref{def:rescale_theta_sigma}, re-scale OU parameters to satisfy the identifiability constraint.}
        \item \textit{Check for block-wise convergence: Let $\bm{\Theta}$ be a vector containing all elements of $\bm{\Lambda}$, $\bm{\Sigma}_u$, $\bm{\Sigma}_{\epsilon}$, $\bm{\theta}$, and $\bm{\sigma}$. Then, calculate $$\delta = \max \Big\{I\big\{|\bm{\Theta}^{(r)} - \bm{\Theta}^{(r-1)}|/\bm{\Theta}^{(r)} < 10^{-6}\big\}, \ I\big\{logL(\bm{\Theta}^{(r)} | \bm{Y}) - logL(\bm{\Theta}^{(r-1)} | \bm{Y}) < 10^{-6}\big\}\Big\}$$ where all operations on $\bm{\Theta}$ are element-wise.}
        \item \textit{Update} $r$: $r = r+1$
    \end{enumerate}
    \item \textit{Estimate Fisher Information-based standard errors from a numerical approximation to the Hessian of the log-likelihood, $\frac{\partial^2}{\partial\Theta \partial\Theta^{\top}}logL(\bm{\Lambda}^{(r)}, \bm{\Sigma}_u^{(r)}, \bm{\Sigma}_{\epsilon}^{(r)}, \bm{\theta}^{(r)} | Y).$}
\end{enumerate}

Note that when estimating standard errors, the parameterization of the likelihood differs slightly: the likelihood now depends on only one of the parameter matrices in the structural submodel, $\bm{\theta}$, and not the other, $\bm{\sigma}$. This change in parameterization is a result of the identifiability constraint that is placed on the stationary variance of the OU process. Since we are no longer conditioning on fixed measurement submodel parameters in this step, we restrict $\bm{\sigma}$ to be a function of $\bm{\theta}$, where this function is derived from the identifiability constraint; thus, the likelihood is not over-parameterized. Standard error estimates for $\bm{\sigma}$ can be calculated via parametric bootstrap. By sampling values of $\bm{\theta}$ from a Normal distribution defined by its point estimate and estimated covariance matrix, bootstrapped values of $\bm{\sigma}$ are calculated as a function of $\bm{\theta}$ and a confidence interval can be estimated based on the empirical distribution. More details on the parameterization of the log-likelihood for standard error estimation are in Section A.6 of the supplementary material. 

To increase the computational efficiency of this estimation algorithm, we (i) leverage the Markov property of the OU process and use the computationally-simpler sparse precision matrix derived in Lemma \ref{def:precision_matrix} rather than the dense covariance matrix, (ii) take advantage of tractable analytic gradients for the measurement submodel given in Lemma \ref{def:fa_gradients}, avoiding the need to calculate computationally expensive numerical gradients, and (iii) implement portions of our algorithm in C++.

\section{Simulation study}\label{s:sim_study}

\subsection{Data generation}

We conduct a simulation study to assess the bias and variance of estimates produced by our model. We assume that there are $K = 4$ longitudinal outcomes recorded over time for $N = 200$ individuals. For individual $i$, these longitudinal outcomes are measured at $n_i$ different occasions where $n_i \sim Uniform(10, 20)$. The gap times between measurement occasions are drawn from a $Uniform(0.1, 2)$ distribution. We consider simulated data in three different settings in which the true bivariate OU process has varying degrees of autocorrelation (see Section A.8 of the supplementary material for details). Using each true OU process, we generate the observed longitudinal outcomes by drawing from $\bm{Y}_i \sim N(0, \bm{\Sigma}_i^*)$ where $\bm{\Sigma}_i^*$ is defined using
\begin{equation}
\begin{aligned}
    \bm{\Lambda} = \begin{bmatrix}
    1.2 & 0 \\
    1.8 & 0 \\
    0 & -0.4 \\
    0 & 2
    \end{bmatrix} \text{,  }
    \bm{\Sigma}_u = \begin{bmatrix}
    1.1 & 0 & 0 & 0 \\
    0 & 1.3 & 0 & 0 \\
    0 & 0 & 1.4 & 0 \\
    0 & 0 & 0 & 0.9
    \end{bmatrix} \text{, and  }
    \bm{\Sigma}_{\epsilon} = \begin{bmatrix}
    0.6 & 0 & 0 & 0 \\
    0 & 0.5 & 0 & 0 \\
    0 & 0 & 0.4 & 0 \\
    0 & 0 & 0 & 0.7
    \end{bmatrix}.
\end{aligned}
\end{equation}

\noindent When fitting this model, we assume that the structural zeros within the loadings matrix and random intercept covariance matrix are known.

Importantly, some of the parameter values used to generate the data are different from the parameters that will be estimated by the model; this difference is a side-effect of the identifiability assumption. While unbiased estimates of $\bm{\Sigma}_u$ and $\bm{\Sigma}_{\epsilon}$ will match the values used in data generation, the values of $\bm{\Lambda}$ and the OU process parameters $\bm{\theta}$ and $\bm{\sigma}$ will differ. As a result of the re-scaling approach for identification described in Section \ref{ss:identification}, the estimated OU process has a stationary variance of 1. The additional variation present in the OU process during data generation must be absorbed by the loadings matrix $\bm{\Lambda}$. Specifically, the data-generating loadings matrix will be re-scaled according to $\bm{\Lambda} \bm{D}$ where $\bm{D}:= \sqrt{diag\{ V(\bm{\theta}, \bm{\sigma})\}}$ and $\bm{V}$ is the stationary variance of the OU process as defined in Lemma \ref{def:precision_matrix}.  $\bm{\Lambda} \bm{D}$ will be estimated by our algorithm.  The data-generating OU parameters $\bm{\theta}$ and $\bm{\sigma}$ will be re-scaled according to scalar constants chosen such that the stationary variance of the re-scaled OU process is equal to 1 via Lemma \ref{def:rescale_theta_sigma}.  True parameter values indicated in the simulation results have all been re-scaled to match the values targeted by our estimation algorithm. In setting 2, the true OU process used to generate data does have a stationary variance equal to 1 and thus the target parameter values do match the data-generating parameter values.

\subsection{Simulation results}\label{ss:sim_results}

\begin{figure}
    \centering
        \includegraphics[width=0.7\linewidth]{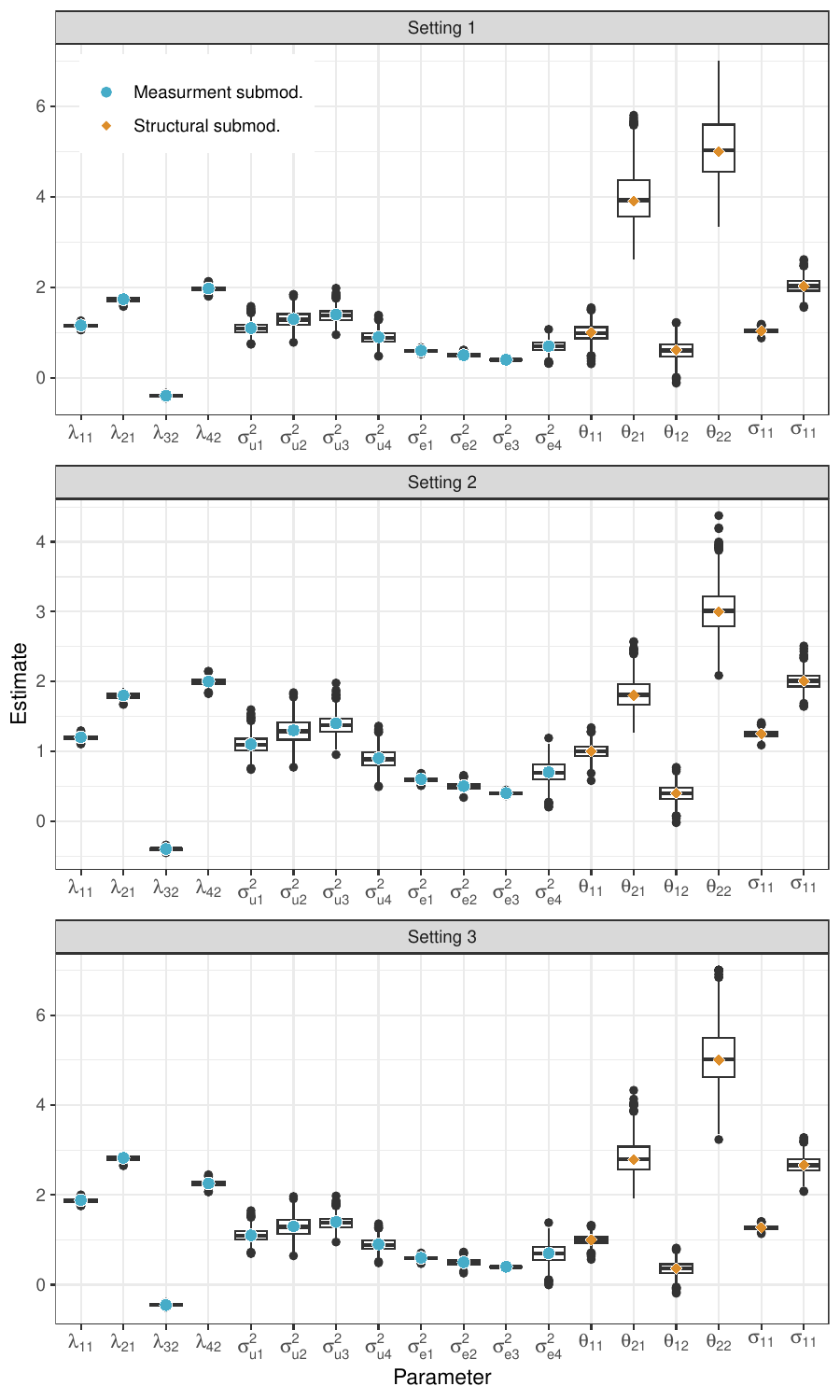}
    \caption{Final parameter estimates from the block coordinate descent algorithm for the three different settings in which the true OU process differs. Point estimates are summarized across the 1000 simulated datasets with box plots and the dots indicate the true (target) parameter values.}\label{fig:point_ests}
\end{figure}

\begin{figure}
    \centering
    \includegraphics[width=0.7\linewidth]{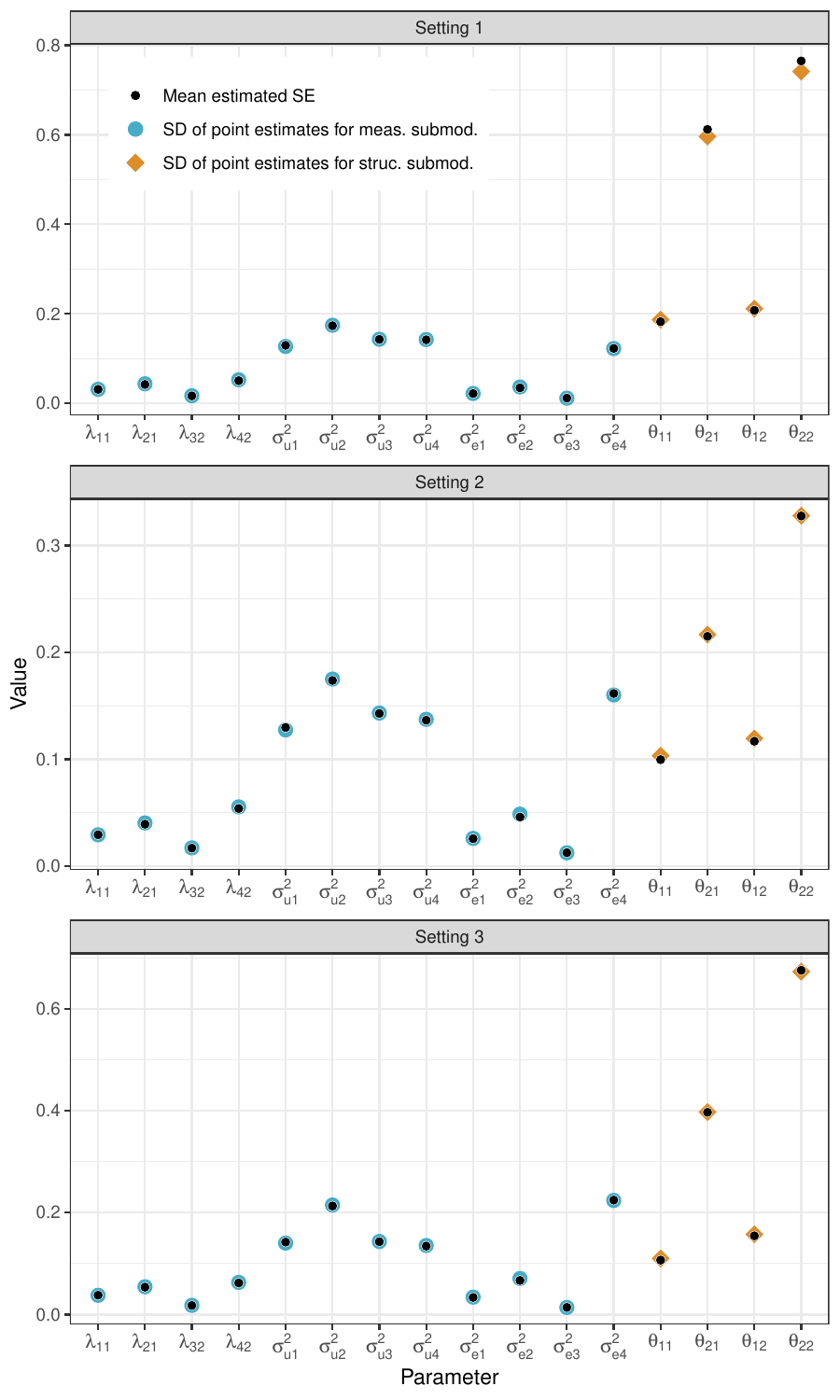}
    \caption{Comparison of estimated standard errors (from Fisher information) and standard deviation of point estimates. The similarity of the standard error estimates and empirical standard deviation suggests that the standard errors are of appropriate size. Note that the standard error estimate for $\sigma_{\epsilon_4}$ is missing for one datasets in Setting 3 (see Section A.9 of the supplementary material for details).}\label{fig:se} 
\end{figure}

In each of the three settings, we generate 1,000 datasets and carry out the estimation algorithm described in Section \ref{ss:est_approach}. Final point estimates are shown in Figure \ref{fig:point_ests} and information-based standard errors are summarized in Figure \ref{fig:se}. In all settings, we consistently recover unbiased estimates of the true values and find that the average of the standard errors is similar to the empirical standard deviation of the point estimates, indicating that confidence intervals will have close to nominal coverage. In a rare case, numerical issues result in a negative variance estimate; this specific case is discussed in Section A.9 of the supplementary material. 

\subsection{Model selection}\label{ss:model_selection}

We next carry out a simulation study in which we evaluate the ability of Akaike information criterion (AIC) and Bayesian information criterion (BIC) to correctly select the true number of latent factors among the fitted models. Formulas for AIC and BIC are given in Section A.7 of the supplementary material. Assuming the same true measurement submodel parameters as before, we now generate data from five different factor models: a one-factor model, a two-factor model with low signal (i.e., high correlation between latent factors), a two-factor model with high signal (i.e., low correlation between latent factors), a three-factor model with low signal, and a three-factor model with high signal.  Data-generating parameter values are given in the Section A.8 of the supplementary material.  For 100 datasets generated from each of these true models, we fit a one-, two-, and three-factor model and compare fit indices.  We do not consider a four-factor model in this simulation study because our data only contain four longitudinal outcomes and so fitting a four-factor model would no longer fall into the dimension-reduction setting in which factor models are generally used.

We present model selection results in Table \ref{tab:mod_selection}.  In both the high and low signal settings, the model with the lowest AIC and BIC most often has the same number of factors as the true model used to generate the data. For models fit to data generated from a true model with three factors, BIC incorrectly selects a model with two factors more often than AIC. This difference make sense given the increased penalty that BIC places on model complexity. For datasets of this size ($N = 200$), estimation becomes more difficult as the number of factors increases and so our algorithm did not converge for a few simulated datasets (see Section A.9 of the supplementary material for details). 

\begin{table}
    \centering
    \includegraphics[width=\linewidth]{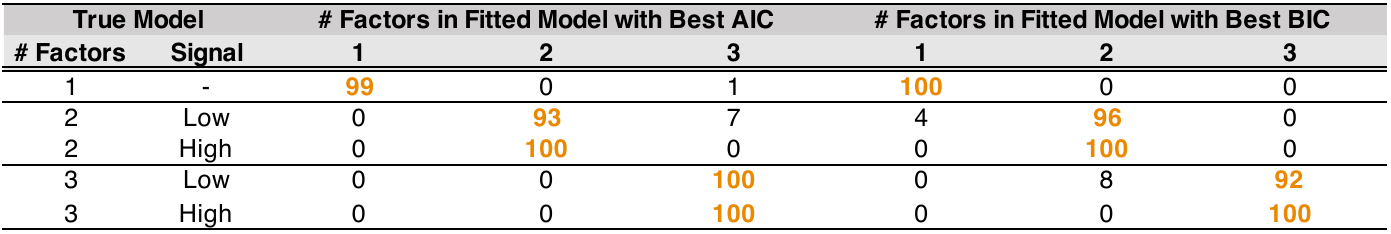}
    \caption{For datasets generated under each true model, we summarize the percent of times that the model-selection metric chose the fitted model with the indicated number of factors. When generating data from models with 2 and 3 factors, we considered two different settings: a high signal setting in which latent factors have lower correlation and a low signal setting in which latent factors have high correlation. The settings in which the fitted model has the same number of factors as the true data-generating model are emphasized with bold orange text. These results are presented for datasets on which the algorithm either converged or reached the maximum number of iterations (200) for all three models. See Section A.9 of the supplementary material for more details.}\label{tab:mod_selection} 
\end{table}

\section{Application to mHealth emotion data}\label{s:data_app}

For illustrative purposes, we apply our method to the data on momentary emotions collected in the mHealth study. Using the OUF model, we summarize the longitudinal responses to 18 emotion-related questions as two latent factors interpreted as positive and negative affect. Positive and negative affect are two distinct-but-correlated emotional states known to be key in understanding smoking habits (e.g., \citet{minami_2014}, \citet{langdon_2016}, \citet{leventhal_2013}, \cite{baker_2004}). The proposed model accounts for both the rapid temporal variation in these states and their correlation over time.  In these data, positive affect was measured by how strongly individuals agreed with feeling happy, joyful, enthusiastic, active, calm, determined, grateful, proud, and attentive; negative affect was measured by how strongly individuals agreed with feeling sad, scared, disgusted, angry, shameful, guilty, irritable, lonely, and nervous.

\begin{figure}
    \centering
    \includegraphics[width=\linewidth]{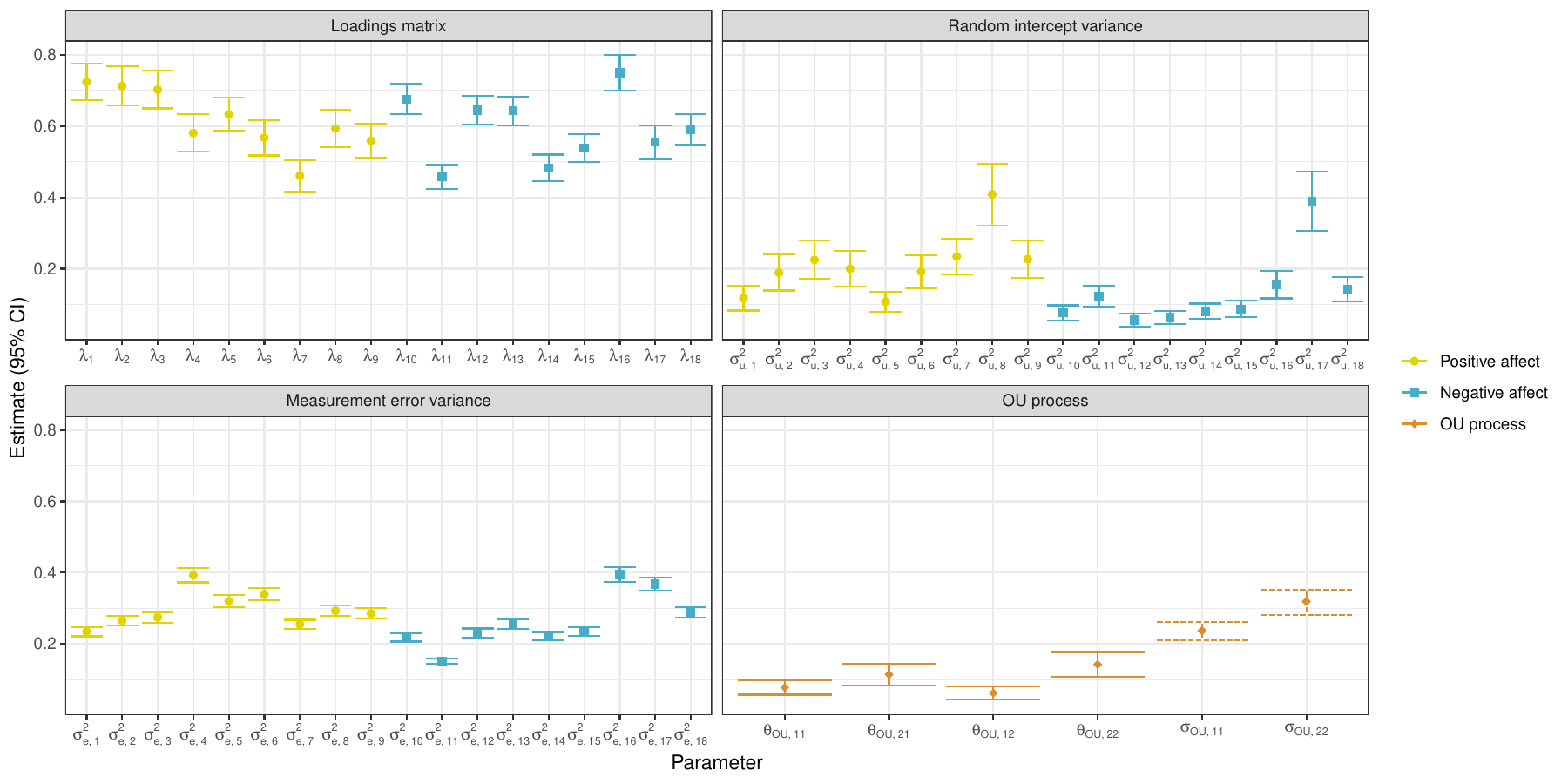}
    \caption{Point estimates and corresponding 95\% confidence intervals (CI) for each of the parameter matrices in our OUF model. Intervals for OU parameters $\sigma_{11}$ and $\sigma_{22}$ are based on a parametric bootstrap. Because we assume structural zeros in the loadings matrix are known, each emotion has only a single loading. Parameter subscripts 1-18 correspond to the emotions as follows: 1 = happy, 2 = joyful, 3 = enthusiastic, 4 = active, 5 = calm, 6 = determined, 7 = grateful, 8 = proud, 9 = attentive, 10 = sad, 11 = scared, 12 = disgusted, 13 = angry, 14 = ashamed, 15 = guilty, 16 = irritable, 17 = lonely, 18 = nervous.} \label{fig:real_dat_results}
\end{figure}

\begin{figure}
    \centering
    \includegraphics[width=0.8\linewidth]{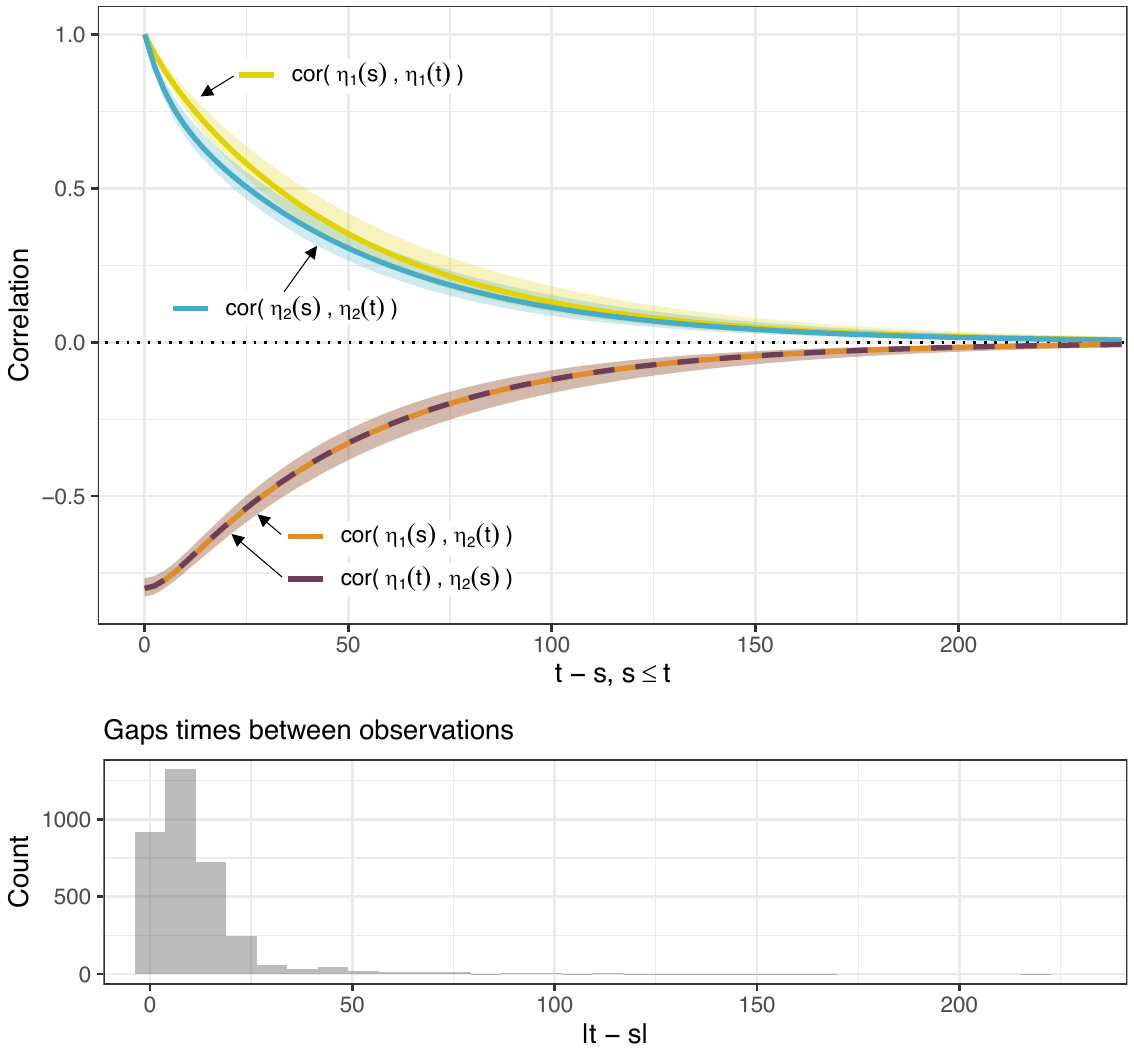}
    \caption{The top panel shows the decay in autocorrelation and cross-correlation between latent factors that represent positive affect ($\eta_1(t)$) and negative affect ($\eta_2(t)$) across increasing gap times, where time is measured in hours. Curves are calculated using OU parameters estimated from emotions measured in the mHealth study. The shaded bands indicate the 2.5th and 97.5th percentiles of a parametric bootstrap. The bottom plot summarizes the distribution of the observed gap times (in hours) between measurements for all individuals in the mHealth study.} \label{fig:real_dat_autocor}
\end{figure}

When applying the OUF model, the assumed structural zeros within the loadings matrix result in positive emotions loading onto one of the latent variables, $\bm{\eta}_1(t)$, and negative emotions loading onto the other, $\bm{\eta}_2(t)$. Point estimates and 95\% confidence intervals are in Figure \ref{fig:real_dat_results}. Measures of happiness, joy, and enthusiasm are most strongly correlated with positive affect and measures of sadness and irritability are most strongly correlated with negative affect. We use the estimated parameters of the OU process to understand the latent dynamics of positive and negative affect by plotting the degree of correlation for these two latent variables across varying time intervals between consecutive observations (see Figure \ref{fig:real_dat_autocor}). We see that positive and negative affect are negatively correlated as expected, and that the correlation between the latent states decays slowly.

We also fit a univariate OUF model and a trivariate OUF model and compare these models to the bivariate OUF model.  In the univariate OUF model, all emotions are assumed to be generated from a single common underlying factor; in the trivariate OUF model, we further divide the positive emotions into two latent factors that we call high arousal positive affect (measured by feeling grateful, proud, enthusiastic, active, determined, attentive) and no-to-low arousal positive affect (measured by feeling calm, happy, and joyful), while the negative emotions are still assumed to be generated from one latent factor.  Coefficient estimates from these fitted models are given in Section B.1 of the supplementary material.

Both AIC and BIC indicate that, of the three models considered, the two factor model fits best: $AIC_{\text{1 factor}} = 123,309$ vs. $AIC_{\text{2 factors}} = 121,069$ vs. $AIC_{\text{3 factors}} = 124,957$ and $BIC_{\text{1 factor}} = 123,791$ vs. $BIC_{\text{2 factor}} = 121,577$ vs. $BIC_{\text{3 factor}} = 125,509$.  We provide more details on the calculation of AIC and BIC in Section B.1 of the supplementary material.  Psychological literature supports our conclusion that two factors represent our data better than one as it suggests that positive and negative affect are not opposites, rather they capture distinct-but-correlated components of psychological state (\citealp*{reich_2003}). The lower AIC and BIC of the two factor model compared to the three factor model suggest that the emotions corresponding to high arousal positive affect and no-to-low arousal positive affect are not different enough to justify the additional complexity of the three factor model given the current data.

\section{Discussion}\label{s:discussion}

We developed a dynamic OUF model that combines a factor model to summarize multivariate observed longitudinal outcomes as lower dimensional latent factors and an OU process to describe the temporal evolution of the latent factors in continuous time. By using the OU process, instead of a discrete time approach such as a VAR process, our model can be applied to irregularly-measured ILD commonly produced by mHealth studies. The OU process captures rapid variations in the correlated latent factors over time, in contrast to a multivariate mixed model that is more suitable for capturing smooth trends over time. To fit our model, we use a block coordinate descent algorithm to directly maximize the log-likelihood of the observed multivariate longitudinal outcome.  We derive both the close-form likelihood of the measured outcome and the sparse precision matrix for the multivariate OU process.  Our block coordinate descent algorithm leverages analytic gradients for a subset of parameters to improve computational efficiency. Finally, we applied our method to study the dynamics of emotions among smokers attempting to quit.

Through the marginal distribution of the multivariate OU process, we parameterize our likelihood in terms of the standard OU drift ($\bm{\theta}$) and volatility ($\bm{\sigma}$) parameters. Having estimates for these parameters enables us to understand the dynamics of the latent factors, including generating new trajectories using the estimated values and examining the decay in the trajectories' correlation over time. Through examination of decay in correlation over time, our method could help inform the design of future studies that aim to collect ILD by providing insight into how frequently the longitudinal outcomes must be measured in order to capture the correlation between them.

In our simulation study in Section \ref{s:sim_study}, we generated data under true OU processes that showed reasonably slow decay in correlation over time given the intervals between measurements. We found that estimation of the OU parameters is difficult if correlation decays quickly relative to gaps between measurements. When longitudinal outcomes are measured frequently enough that correlation between consecutive measurements is captured, our method consistently returns unbiased estimates of the OU process parameters. If this method were applied to data in which longitudinal outcomes are not measured often enough to capture the correlation, estimation would be difficult. Like all statistical methods, when enough signal exists in the data, our method works well. It does, however, require studies to be designed such that longitudinal outcomes are measured with sufficient frequency that the correlation between consecutive measurements is captured.

Although we use the sparse OU precision matrix, leverage the availability of analytic gradients for the measurement submodel parameters, and implement a portion of our algorithm in C++, the computation time of our estimation algorithm increases rapidly as the number of longitudinal outcomes increases. We successfully fit our model to a dataset containing 18 longitudinal outcomes but this does require approximately 27 hours. In order to make application of our model to datasets with larger numbers of longitudinal outcomes feasible, computational efficiency must be increased.  However, our proposed marginal likelihood-based method has substantial computational benefits when compared to alternative methods.  In comparison to the Bayesian approach proposed for fitting a similar model in \cite{tran_2021b}, our approach requires less computation time.  In a simulation study with $K = 4$ longitudinal outcomes measured at 10-20 occasions on $N = 200$ individuals, we found that estimation via our block coordinate descent algorithm required approximately 5\% of the time required by the Bayesian approach proposed in \cite{tran_2021b} given the same computing resources.  More details on this comparison are given in Section C.2 of the supplementary material.

Finally, the mHealth dataset to which we applied our method also contains information on demographic characteristics and on the timing of cigarette use. Including baseline covariates in either the measurement or structural submodel would be a useful extension. In behavioral science, specific emotional states, such as negative affect or craving, are expected to be correlated with cigarette use and so future work could involve combining our OUF model with a submodel for event-time outcomes. Our model could also be modified to account for treatment or for drift in the OU process to better capture the dynamics of the latent processes after a key event such as smoking cessation or relapse.

\section*{Acknowledgements}

This work was supported by the National Institutes of Health [grant numbers F31DA057048, P30CA042014, P50DA054039, R01DA039901, R01MD010362, T32CA083654, U01CA229437]; and the Huntsman Cancer Foundation. The National Institutes of Health had no role in the study design, collection, analysis or interpretation of the data, writing the manuscript, or the decision to submit the paper for publication. The content is solely the responsibility of the authors and does not necessarily represent the official views of the National Institutes of Health or the Huntsman Cancer Foundation. This work is not peer reviewed. The authors declare no conflicts of interest.

\section*{Supporting Information}

Supplementary material is available with this paper.  Example code and simulated data are available on Github at https://github.com/madelineabbott/OUF.

\vspace*{-8pt}

\bibliographystyle{biom} 
\bibliography{references.bib}

\label{lastpage}

\end{document}


\maketitle

\renewcommand{\figurename}{Supplementary Figure}
\renewcommand{\tablename}{Supplementary Table}
\renewcommand{\thesection}{A.\arabic{section}}

\setcounter{section}{0}
\section*{Section A}

\section{Derivation of the analytic form of the conditional covariance function of the OU process}\label{a:integral_free}

Assume $\eta(t)$ is a $p$-dimensional Ornstein-Uhlenbeck (OU) stochastic process with a marginal mean of $0$.  From \cite{vatiwutipong_2019}, if we assume that the initial state $\eta(t_0 = 0)$ is known, then the cross-covariance function of the OU process at times $s$ and $t$ is 

\begin{equation*}
\begin{aligned}
    Cov\{\eta(s), \eta(t) | \eta(t_0 = 0) \} = \int_{0}^{\min(s,t)} e^{-\theta(s-u)} \sigma \sigma^{\top} e^{-\theta^{\top}(t-u)}du
\end{aligned}
\end{equation*}

where $e^A$ is the matrix exponential. Note that we can assume that $t_0 = 0$ without loss of generality because this stochastic process is stationary.  Using the identity for matrices $A$, $B$, and $C$ that $vec(ABC) = (C^{\top} \otimes A) vec(B)$, we can re-write the vectorized version of the cross-covariance function as
\begin{equation*}
\begin{aligned}
    vec\{Cov\{\eta(s), \eta(t) | \eta(t_0) \}\} = \int_{0}^{\min(s,t)} e^{-\theta(t-u)} \otimes e^{-\theta(s-u)} \ du \  vec\{\sigma \sigma^{\top}\}
\end{aligned}
\end{equation*}

We can also use the identity that $e^A \otimes e^B = e^{A \oplus B}$, so
\begin{equation}
\begin{aligned}
\label{eq:xcov_int_cond}
    vec\{Cov\{\eta(s), \eta(t) | \eta(t_0) \}\} = \int_{0}^{\min(s,t)} e^{[-\theta(t-u)] \oplus [-\theta(s-u)]} du \ vec\{\sigma \sigma^{\top}\}
\end{aligned}
\end{equation}

Next, we simplify Equation \ref{eq:xcov_int_cond} by pulling all the $u$'s into a single term. For now, focus on the term in the exponential:
\begin{equation*}
\begin{aligned}
    \left[-\theta(t-u)\right] \oplus \left[-\theta(s-u)\right] &\overset{(a)}{=} -\theta(t-u) \otimes I + I \otimes (-\theta(s-u)) \\
    &= -t(\theta \otimes I) + u(\theta \otimes I + I \otimes \theta) - s(I \otimes \theta) \\
    &= -(t\theta \oplus s\theta) + u (\theta \oplus \theta)
\end{aligned}
\end{equation*}

where equality $(a)$ is by the definition of the Kronecker sum; $A \oplus B = A \otimes I_B + I_A \otimes A$, where $I_A$ and $I_B$ are identity matrices with dimensions of $A$ and $B$, respectively. Now, substituting this new term back into the exponential term in Equation \ref{eq:xcov_int_cond}, we get
\begin{equation}
\begin{aligned}
\label{eq:exp_term}
    e^{[-\theta(t-u)] \oplus [-\theta(s-u)]} = e^{-(t\theta \oplus s\theta) + u (\theta \oplus \theta)}
\end{aligned}
\end{equation}

We can simplify this further using the identity $e^{A + B} = e^A e^B$ if $A$ and $B$ commute. Letting $A = (t\theta) \oplus (s\theta)$ and $B = (\theta \oplus \theta)$, we first show that these terms commute:
\begin{equation*}
\begin{aligned}
    A \cdot B = &\left[ (t \theta) \oplus (s \theta) \right] \cdot \left[ \theta \oplus \theta \right] \\
    =&\left[ t \theta \otimes I + I \otimes s \theta \right] \cdot \left[ \theta \otimes I + I \otimes \theta \right] \\
    =& (t\theta \otimes I) (\theta \otimes I) + (t \theta \otimes I) (I \otimes \theta)  + (I \otimes s \theta) (\theta \otimes I) + (I \otimes s \theta) (I \otimes \theta) \\
    =& (t \theta \otimes I) (\theta \otimes I) + (I \otimes \theta) (t \theta \otimes I)  + (\theta \otimes I) (I \otimes s \theta) + (I \otimes s \theta) (I \otimes \theta) \\
    =&(\theta \otimes I) \left[ (t \theta \otimes I) + (I \otimes s \theta) \right] + (I \otimes \theta) \left[ (t \theta \otimes I)  + (I \otimes s \theta) \right] \\
    =&\left[ (\theta \otimes I) + (I \otimes \theta) \right] \cdot \left[ (t \theta \otimes I)  + (I \otimes s \theta) \right] \\
    =&\left[ (\theta \oplus \theta) \right] \cdot \left[ (t \theta \oplus s \theta) \right] \\
\end{aligned}
\end{equation*}

where line 4 uses the mixed-product property of the Kronecker product.

Referring back to Equation \ref{eq:exp_term}, we now have
\begin{equation*}
\begin{aligned}
    e^{-(t\theta \oplus s\theta) + u (\theta \oplus \theta)} = e^{-(t\theta \oplus s\theta)} e^{u(\theta \oplus \theta)}
\end{aligned}
\end{equation*}

We can substitute this term into Equation \ref{eq:xcov_int_cond} to get
\begin{equation*}
\begin{aligned}
    vec\{Cov\{\eta(s), \eta(t) | \eta(t_0 = 0) \}\} &= \int_{0}^{\min(s,t)} e^{-(t\theta \oplus s\theta)} e^{u(\theta \oplus \theta)} du \ vec\{\sigma \sigma^{\top}\} \\
    &= \int_{0}^{\min(s,t)} e^{u(\theta \oplus \theta)} du \ e^{-(t\theta \oplus s\theta)} vec\{\sigma \sigma^{\top}\}
\end{aligned}
\end{equation*}

Now that we have rewritten the conditional cross-covariance function in this form, the only term that we need to integrate is $e^{u(\theta \oplus \theta)}$. We find
\begin{equation*}
\begin{aligned}
    \int_0^{min(s,t)} e^{u(\theta \oplus \theta)} du = (\theta \oplus \theta)^{-1} \big[ e^{min(s,t)(\theta \oplus \theta)} - I \big]
\end{aligned}
\end{equation*}

We now have an integral-free analytic form of the conditional cross-covariance function:
\begin{equation*} 
\begin{aligned}
    vec\{Cov\{\eta(s), \eta(t) | \eta(t_0 = 0) \}\} &= (\theta \oplus \theta)^{-1} \big[ e^{min(s,t) (\theta \oplus \theta)} - I \big] e^{-(t \theta \oplus s \theta)} vec\{ \sigma \sigma^{\top} \}
\end{aligned}
\end{equation*}

Note that if $s = t$, then the conditional cross-covariance function simplifies to the conditional covariance function given in \citet{vatiwutipong_2019}.

\section{Derivation of the analytic form of the marginal covariance function of the OU process}\label{a:marginal_xcov}

The analytic form of the conditional covariance function given in Lemma 1 (in the main text) is based on the assumption that the initial state $\eta(t_0)$, with $t_0 = 0$ is \textit{known}.  We now derive the analytic form of the unconditional cross-covariance function that accounts for the additional uncertainty of an unknown initial state.  From \citet{vatiwutipong_2019}, if $\eta(t_0)$, with $t_0 = 0$, is known, then
\begin{equation*}
\begin{aligned}
    \mathbb{E}\{ \eta(t) | \eta(t_0) \} = e^{-\theta t} \eta(t_0)
\end{aligned}
\end{equation*}

Assuming that $s \le t$, from Lemma 1, we have
\begin{equation*}
\begin{aligned}
    Cov\{ \eta(s), \eta(t) | \eta(t_0) \} = vec^{-1}\Big\{(\theta \oplus \theta)^{-1}\Big[ e^{(\theta \oplus \theta) s} - I \Big] e^{-(\theta t \oplus \theta s)} vec\{\sigma \sigma^{\top} \} \Big\}
\end{aligned}
\end{equation*}


If $\eta(t_0)$ is \textit{unknown} and $t_0 = 0$, then using the Law of Total Covariance we can calculate
\begin{equation*}
\begin{aligned}
    Cov\{ \eta(s), \eta(t)\} &= \mathbb{E}\big\{ Cov\big( \eta(s), \eta(t) | \eta(t_0) \big) \big\} + Cov\big\{ \mathbb{E}\big( \eta(s) | \eta(t_0) \big), \mathbb{E}\big( \eta(t) | \eta(t_0) \big) \big\} \\
    &=  vec^{-1}\Big\{(\theta \oplus \theta)^{-1}\Big[ e^{(\theta \oplus \theta) s} - I \Big] e^{-(\theta t \oplus \theta s)} vec\{\sigma \sigma^{\top} \} \Big\} \\ & \ \ + Cov\big\{ e^{-\theta s} \eta(t_0), e^{-\theta t} \eta(t_0) \big\} \\
    &= vec^{-1}\Big\{(\theta \oplus \theta)^{-1}\Big[ e^{(\theta \oplus \theta) s} - I \Big] e^{-(\theta t \oplus \theta s)} vec\{\sigma \sigma^{\top} \} \Big\} \\ & \ \ + e^{-\theta s} Var\big\{  \eta(t_0)\big\} [e^{-\theta t}]^{\top}
\end{aligned}
\end{equation*}

If we assume that $\eta(t_0)$ is drawn from the stationary distribution, then $Var(\eta(t_0)) = vec^{-1}\big\{ (\theta \oplus \theta)^{-1} vec\{ \sigma \sigma^{\top} \} \big\}$. Then, we have 
\begin{equation*}
\begin{aligned}
    Cov\{ \eta(s), \eta(t)\} =& vec^{-1}\Big\{(\theta \oplus \theta)^{-1}\Big[ e^{(\theta \oplus \theta) s} - I \Big] e^{-(\theta t \oplus \theta s)} vec\{\sigma \sigma^{\top} \} \Big\} \\ &+ e^{-\theta s} vec^{-1}\big\{ (\theta \oplus \theta)^{-1} vec\{\sigma \sigma^{\top} \} \big\}  [e^{-\theta t}]^{\top}
\end{aligned}
\end{equation*}

Now we simplify this function. Consider the terms involving $\theta$ in the first term of the sum,
\begin{equation*}
\begin{aligned}
    (\theta \oplus \theta)^{-1}\Big[ e^{(\theta \oplus \theta) s} - I \Big] e^{-(\theta t \oplus \theta s)}
\end{aligned}
\end{equation*}

We can simplify this expression using the fact that $e^A e^B = e^B e^A$ in our setting. This property means that both
\begin{equation}
\begin{aligned}
\label{eq:commute1}
    (\theta \oplus \theta)^{-1}\big[ e^{s (\theta \oplus \theta)} - I \big] e^{-(t \theta \oplus s \theta)} = e^{-(t \theta \oplus s \theta)} (\theta \oplus \theta)^{-1}\big[ e^{s (\theta \oplus \theta)} - I \big] 
\end{aligned}
\end{equation}

and

\begin{equation}
\begin{aligned}
\label{eq:commute2}
    (\theta \oplus \theta)^{-1}\big[ e^{s (\theta \oplus \theta)} - I \big] e^{- (t \theta \oplus s \theta)} = (\theta \oplus \theta)^{-1} e^{-(t \theta \oplus s \theta)} \big[ e^{s (\theta \oplus \theta)} - I \big]
\end{aligned}
\end{equation}

Setting Equations \ref{eq:commute1} and \ref{eq:commute2} equal and cancelling the final term implies that 
\begin{equation*}
\begin{aligned}
    e^{-(t \theta \oplus s \theta)} (\theta \oplus \theta)^{-1} = (\theta \oplus \theta)^{-1} e^{-(t \theta \oplus s \theta)}
\end{aligned}
\end{equation*}

We will use this proof of the commutative property later and now return to our expression for the unconditional cross-covariance function, $Cov\{ \eta(s), \eta(t) \}$,
\begin{equation}
\begin{aligned}
\label{eq:xcov_intermed}
    Cov\{ \eta(s), \eta(t)\} =& vec^{-1}\Big\{(\theta \oplus \theta)^{-1}\Big[ e^{(\theta \oplus \theta) s} - I \Big] e^{-(\theta t \oplus \theta s)} vec\{\sigma \sigma^{\top} \} \Big\} \\&+ e^{-\theta s} vec^{-1}\big\{ (\theta \oplus \theta)^{-1} vec\{\sigma \sigma^{\top} \} \big\}  [e^{-\theta t}]^{\top}
\end{aligned}
\end{equation}

Consider the second term in the sum,
\begin{equation*}
\begin{aligned}
    e^{-\theta s} vec^{-1}\big\{ (\theta \oplus \theta)^{-1} vec\{\sigma \sigma^{\top} \} \big\}  [e^{-\theta t}]^{\top}
\end{aligned}
\end{equation*}

By applying the identity $vec(ABC) = (C^{\top} \otimes A) vec(B)$, we can rewrite the vectorized form of the expression as
\begin{equation*}
\begin{aligned}
    vec\big\{e^{-\theta s} vec^{-1}\big\{ (\theta \oplus \theta)^{-1} vec\{\sigma \sigma^{\top} \} \big\}  [e^{-\theta t}]^{\top} \big\} &= e^{-\theta t} \otimes e^{-\theta s} vec\big\{ vec^{-1}\{ (\theta \oplus \theta)^{-1} vec\{ \sigma \sigma^{\top} \} \big\} \\
    &= e^{-\theta t} \otimes e^{-\theta s} (\theta \oplus \theta)^{-1} vec\{ \sigma \sigma^{\top} \} \\
    &= e^{-(\theta t \oplus \theta s)} (\theta \oplus \theta)^{-1} vec\{ \sigma \sigma^{\top} \}
\end{aligned}
\end{equation*}

Reversing the vectorization operation and applying the commutative property, we then get
\begin{equation*}
\begin{aligned}
    e^{-\theta s} vec^{-1}\big\{ (\theta \oplus \theta)^{-1} vec\{\sigma \sigma^{\top} \} \big\}  [e^{-\theta t}]^{\top} &= vec^{-1} \big\{ e^{-(\theta t \oplus \theta s)} (\theta \oplus \theta)^{-1} vec\{ \sigma \sigma^{\top} \} \big\} \\
    &= vec^{-1} \big\{ (\theta \oplus \theta)^{-1} e^{-(\theta t \oplus \theta s)} vec\{ \sigma \sigma^{\top} \} \big\}
\end{aligned}
\end{equation*}

Plugging the term above into the second term of Equation \ref{eq:xcov_intermed}, the cross-covariance function becomes
\begin{equation}
\begin{aligned}
\label{eq:xcov}
    Cov\{ \eta(s), \eta(t) \} &= vec^{-1} \Big\{ (\theta \oplus \theta)^{-1} \Big[ e^{(\theta \oplus \theta) s} e^{-(\theta t \oplus \theta s)} - e^{-(\theta t \oplus \theta s)} \Big] vec\{ \sigma \sigma^{\top} \} \Big\} \\ & \ \ \ + vec^{-1} \Big\{ (\theta \oplus \theta)^{-1} e^{-(\theta t \oplus \theta s)} vec\{ \sigma \sigma^{\top} \} \Big\} \\
    &= vec^{-1} \Big\{ (\theta \oplus \theta)^{-1} e^{(\theta \oplus \theta)s} e^{-(\theta t \oplus \theta s)} vec\{ \sigma \sigma^{\top}\} - (\theta \oplus \theta)^{-1} e^{-(\theta t \oplus \theta s)} vec\{ \sigma \sigma^{\top} \} \Big\} \\ & \ \ \ + vec^{-1} \Big\{ (\theta \oplus \theta)^{-1} e^{-(\theta t \oplus \theta s)} vec\{ \sigma \sigma^{\top} \} \Big\}\\
    &= vec^{-1} \Big\{ (\theta \oplus \theta)^{-1} \big[ e^{(\theta \oplus \theta) s} e^{-(\theta t \oplus \theta s)} \big] vec\{ \sigma \sigma^{\top} \} \Big\} \\
    &= vec^{-1} \Big\{ (\theta \oplus \theta)^{-1} \big[ e^{(\theta \oplus \theta)s - (\theta t \oplus \theta s)} \big] vec\{\sigma \sigma^{\top}\} \Big\}
\end{aligned}
\end{equation}

Equation \ref{eq:xcov} is the marginal cross-covariance function of the OU process when the initial state at time $t_0 = 0$ is \textit{unknown}.

\section{Derivation of the precision matrix for the OU process}
\label{a:precision}

We derive the sparse precision matrix for the multivariate OU process assuming an unknown initial state. This sparsity results from the Markov property.  We use $\Omega$ to represent the precision matrix and $\Psi$ for the covariance matrix.

\hspace{2em}First, we start in the simplest setting in which we assume a stationary univariate OU process with evenly spaced measurement occasions. The spacing of the measurement times is given by $|t_j - t_{j-1}| =: d > 0$. The covariance matrix takes the form,
\begin{equation*}
\begin{aligned}
    \Psi = \frac{\sigma^2}{2 \theta} \begin{bmatrix} 1 & e^{-\theta d} & \dots & e^{-\theta(n-2)\cdot d} & e^{-\theta(n-1)\cdot d} \\
    e^{-\theta d} & 1 & \dots & e^{-\theta(n-3)\cdot d} & e^{-\theta(n-2)\cdot d} \\ \vdots & \vdots & \ddots & \vdots & \vdots \\
    e^{-\theta(n-2)\cdot d} & e^{-\theta(n-3)\cdot d} & \dots & 1 & e^{-\theta d} \\ e^{-\theta(n-1)\cdot d} & e^{-\theta(n-2)\cdot d} & \dots & e^{-\theta d} & 1 \end{bmatrix}
\end{aligned}
\end{equation*}

We know that the univariate OU process is equal to the AR(1) process when measurements are evenly spaced, so the OU process precision matrix (assuming evenly spaced measurements) can be expressed as
\begin{equation*}
    \Omega = \frac{2\theta}{\sigma^2}\frac{1}{1 - e^{-2\theta d}} \begin{bmatrix} 1 & -e^{-\theta d} & \dots & 0 & 0 \\
    -e^{-\theta d} & 1 + e^{-2\theta d} & \dots & 0 & 0 \\
    \vdots & \vdots & \ddots & \vdots & \vdots \\
    0 & 0 & \dots & 1 + e^{2 \theta d} & -e^{-\theta d} \\
    0 & 0 & \dots & -e^{-\theta d} & 1\end{bmatrix}
\end{equation*}

Now, consider a more general setting in which measurements do not necessarily occur at evenly spaced intervals. Assume that $t_1 < t_2 < \dots < t_{n-1} < t_n$. Then, the covariance matrix takes the form
\begin{equation*}
    \Psi = \frac{\sigma^2}{2 \theta} \begin{bmatrix}
    1 & e^{-\theta |t_2 - t_1|} & \dots & e^{-\theta|t_{n-1} - t_1|} & e^{-\theta|t_n - t_1|} \\
    e^{-\theta |t_2 - t_1|} & 1 & \dots & e^{-\theta|t_{n-1} - t_2|} & e^{-\theta|t_n - t_2|} \\
    \vdots & \vdots & \ddots & \vdots & \vdots \\
    e^{-\theta|t_{n-1} - t_1|} & e^{-\theta|t_{n-1} - t_2|} & \dots & 1 & e^{-\theta|t_{n-1} - t_n|} \\ e^{-\theta|t_n - t_1|} & e^{-\theta|t_n - t_2|} & \dots & e^{-\theta|t_n - t_{n-1}|} & 1 \end{bmatrix}
\end{equation*}

and the precision matrix can be expressed as
\begin{equation*}
\footnotesize
    \Omega = \frac{2\theta}{\sigma^2} \begin{bmatrix} \frac{1}{1 - e^{-2\theta |t_2 - t_1|}} & -\frac{e^{-\theta |t_2 - t_1|}}{1 - e^{-2\theta |t_2 - t_1|}} & \dots & 0 & 0 \\
    -\frac{e^{-\theta |t_2 - t_1|}}{1 - e^{-2\theta |t_2 - t_1|}} & \frac{1 - e^{-2\theta |t_3 - t_1|}}{(1 - e^{-2\theta |t_2 - t_1|})(1 - e^{-2\theta | t_3 - t_2|})} & \dots & 0 & 0 \\ \vdots & \vdots & \ddots & \vdots & \vdots \\
    0 & 0 & \dots & \frac{1 - e^{-2\theta |t_n - t_{n-2}|}}{(1 - e^{-2\theta |t_2 - t_1|})(1 - e^{-2\theta|t_3 - t_2|})} & -\frac{e^{-2\theta|t_n - t_{n-1}|}}{1 - e^{-2\theta|t_n - t_{n-1}|}} \\
    0 & 0 & \dots & -\frac{e^{-2\theta|t_n - t_{n-1}}}{1 - e^{-2\theta|t_n - t_{n-1}|}} & \frac{1}{1 - e^{-2\theta |t_n - t_{n-1}|}} \end{bmatrix}
\end{equation*}

Next, we move from the one-dimensional case to the two-dimensional case. We start by re-arranging the terms in the definition of the cross-covariance function for the bivariate OU process.
\begin{equation*}
    \begin{aligned}
        Cov\{ \eta(s), \eta(t) \} &= vec^{-1}\big\{ (\theta \oplus \theta)^{-1} e^{s \wedge t (\theta \oplus \theta) - (\theta t) \oplus (\theta s)} vec(\sigma \sigma^{\top})\big\} \\
        & \overset{(a)}{=} vec^{-1}\big\{ e^{s \wedge t (\theta \oplus \theta)} e^{-(\theta t) \oplus (\theta s)} (\theta \oplus \theta)^{-1} vec(\sigma \sigma^{\top}) \big\} \\
        &= vec^{-1}\big\{ \big[ e^{s \wedge t \theta} \otimes e^{s \wedge t \theta} \big] \big[ e^{-\theta t} \otimes e^{-\theta s} \big] (\theta \oplus \theta)^{-1} vec(\sigma \sigma^{\top}) \big\} \\
        &= vec^{-1}\big\{ \big[ e^{s \wedge t \theta} e^{-\theta t}\big] \otimes\big[ e^{s \wedge t \theta} e^{-\theta s} \big] (\theta \oplus \theta)^{-1} vec(\sigma \sigma^{\top}) \big\} \\
        &= vec^{-1}\big\{ \big[ e^{-\theta(t - s \wedge t)}\big] \otimes \big[ e^{-\theta (s - s \wedge t)} \big] (\theta \oplus \theta)^{-1} vec(\sigma \sigma^{\top}) \big\} \\
        &\overset{(b)}{=} vec^{-1}\big\{ \big[ e^{-\theta(t - s \wedge t)}\big] \otimes I (\theta \oplus \theta)^{-1} vec(\sigma \sigma^{\top}) \big\} \\
        &= vec^{-1}\big\{(\theta \oplus \theta)^{-1} vec(\sigma \sigma^{\top}) \big\} e^{-\theta^{\top}|t-s|} \\
        &:= V \cdot e^{-\theta^{\top}|t-s|}
    \end{aligned}
\end{equation*}

where equality (a) is because these terms commute and equality (b) holds when we assume that $min(s,t) = s$. We can make this assumption without loss of generality because the matrices are symmetric. When $min(s,t) = t$, $Cov\{ \eta(s), \eta(t) \} = e^{-\theta |t - s|}V^{\top} = e^{-\theta|t-s|}V$. 
Then, the covariance matrix is given by

\begin{equation*}
    \Psi =\begin{bmatrix}
    V & V e^{-\theta^{\top} |t_2 - t_1|} & \dots & V e^{-\theta^{\top}|t_{n-1} - t_1|} & V e^{-\theta^{\top} |t_n - t_1|} \\
    e^{-\theta |t_2 - t_1|} V & V & \dots & V e^{-\theta^{\top}|t_{n-1} - t_2|} & V e^{-\theta^{\top} |t_n - t_2|} \\
    \vdots & \vdots & \ddots & \vdots & \vdots \\
    e^{-\theta|t_{n-1} - t_1|} V & e^{-\theta|t_{n-1} - t_2|} V & \dots & V & V e^{-\theta^{\top}|t_{n-1} - t_n|} \\ e^{-\theta|t_n - t_1|} V & e^{-\theta|t_n - t_2|}  V & \dots & e^{-\theta|t_n - t_{n-1}|} V & V \end{bmatrix}
\end{equation*}

By the definition of the OU process, we know that the precision matrix, $\Omega = \Psi^{-1}$, is block tri-diagonal. We start by solving for two blocks, $\Omega_{11}$ and $\Omega_{12}$. We assume that $\Omega_{11} = A^{-1}$ and $\Omega_{12} = A^{-1}B$, based on the form of the precision matrix in the case of the univariate OU process.  Based on patterns seen when multiplying the AR(1) precision and covariance matrices, we assume that, for the OU process, the first row of blocks in the precision matrix, $[ \Omega_{11}, \Omega_{12}, 0, \dots, 0 ]$ times the second column of blocks in the covariance matrix, $[ V e^{-\theta^{\top}(t_2 - t_1)}, V, \dots ]^{\top}$, is equal to 0.  So,
\begin{equation*}
\begin{aligned}
    0 &= \Omega_{11} V e^{-\theta^{\top}(t_2 - t_1)} + \Omega_{12} V \\
    \Longrightarrow 0 &= A^{-1} V e^{-\theta^{\top}(t_2 - t_1)} + A^{-1} B V \\
    \Longrightarrow 0 &= V e^{-\theta^{\top}(t_2 - t_1)} + BV \\
    \Longrightarrow BV &= - V e^{-\theta^{\top}(t_2 - t_1)} \\
    \Longrightarrow B &= -V e^{-\theta^{\top}(t_2 - t_1)} V^{-1}
\end{aligned}
\end{equation*}

By similar logic, the first row of blocks in the precision matrix times the first column of blocks in the covariance matrix is equal to the identity matrix.  So,
\begin{equation*}
\begin{aligned}
    I &= \Omega_{11} V + \Omega_{12} e^{-\theta(t_2 - t_1)} V \\
    \Longrightarrow I &= A^{-1} V + A^{-1} B e^{-\theta(t_2 - t_1)} V \\
    \Longrightarrow A &= V + B e^{-\theta(t_2 - t_1)} V
\end{aligned}
\end{equation*}

We know that $B = -V e^{-\theta^{\top}(t_2 - t_1)} V^{-1}$ so
\begin{equation*}
\begin{aligned}
    A &= V - V e^{-\theta^{\top} (t_2 - t_1)} V^{-1} e^{-\theta (t_2 - t_1)} V \\
    \Longrightarrow A^{-1} &= [V - V e^{-\theta^{\top} (t_2 - t_1)} V^{-1} e^{-\theta (t_2 - t_1)} V]^{-1}
\end{aligned}
\end{equation*}

Now we have
\begin{equation*}
\begin{aligned}
    \Omega_{11} &= \big[ V - V e^{-\theta^{\top} (t_2 - t_1)} V^{-1} e^{-\theta (t_2 - t_1)} V \big]^{-1} \\
    \Omega_{12} &= -\big[ V - V e^{-\theta^{\top} (t_2 - t_1)} V^{-1} e^{-\theta (t_2 - t_1)} V \big]^{-1} V e^{-\theta^{\top} (t_2 - t_1)} V^{-1}
\end{aligned}
\end{equation*}

\vspace{0.5cm}

Continuing with this logic, we can check the first row of blocks in $\Omega$ against all other columns of $\Psi$ and see that
\begin{equation*}
\begin{aligned}
    0 &= \Omega_{11} V e^{-\theta^{\top}(t_k - t_1)} + \Omega_{12} V e^{-\theta^{\top}(t_k - t_2)} \\
    &= A^{-1} V e^{-\theta^{\top}(t_k - t_1)} + A^{-1} B V e^{-\theta^{\top}(t_k - t_2)} \\
    &= V e^{-\theta^{\top}(t_k - t_1)} + B V e^{-\theta^{\top}(t_k - t_2)} \\
    &= V e^{-\theta^{\top}(t_k - t_1)} - V e^{-\theta^{\top} (t_2 - t_1)} V^{-1} V e^{-\theta^{\top}(t_k - t_2)} \\
    &= V e^{-\theta^{\top}(t_k - t_1)} - V e^{-\theta^{\top} (t_2 - t_1)} e^{-\theta^{\top}(t_k - t_2)} \\
    &= V e^{-\theta^{\top}(t_k - t_1)} - V e^{-\theta^{\top} (t_k - t_1)} \\
    &= 0
\end{aligned}
\end{equation*}

Now we move to the second row of blocks in $\Omega$.  Because $\Omega = \Omega^{\top}$, we also know that $\Omega_{21} = \Omega^{\top}_{12}$.  This symmetry means that we only need to derive $\Omega_{22}$ and $\Omega_{23}$.  Based on previous results, we have
\begin{equation*}
\begin{aligned}
    \Omega_{23} = - \Big[ V - V e^{-\theta^{\top} (t_3 - t_2)} V^{-1} e^{-\theta (t_3 - t_2)} V \Big]^{-1} V e^{-\theta^{\top} (t_3 - t_2)} V^{-1}
\end{aligned}
\end{equation*}

Then we find the form of $\Omega_{22}$ by once again using the same logic to say that the second row of blocks in $\Omega$ times the second column of blocks in $\Psi$ will be equal to an identity matrix:
\begin{equation*}
\begin{aligned}
    I &= \Omega_{21} V e^{-\theta^{\top}(t_2 - t_1)} + \Omega_{22} V + \Omega_{23} e^{-\theta (t_3 - t_2)} V \\
    \Rightarrow V^{-1} &= \Omega_{21} V e^{-\theta^{\top} (t_2 - t_1)} V^{-1} + \Omega_{22} + \Omega_{23} e^{-\theta (t_3 - t_2)} \\
    \Rightarrow \Omega_{22} &= V^{-1} + V^{-1} e^{-\theta (t_2 - t_1)} V \left[ V - V e^{-\theta^{\top}(t_2 - t_1)} V^{-1} e^{-\theta(t_2 - t_1)} V \right]^{-1^{\top}} V e^{-\theta^{\top} (t_2 - t_1)} V^{-1} \\ \ \ \ \ &+ \left[ V - V e^{-\theta^{\top}(t_3 - t_2)} V^{-1} e^{-\theta(t_3 - t_2)} V \right]^{-1} V e^{-\theta^{\top}(t_3 - t_2)} V^{-1} e^{-\theta (t_3 - t_2)}
\end{aligned}
\end{equation*}

The final terms are then given by:
\begin{equation*}
\begin{aligned}
    I &= \Omega_{n,n-1} V e^{-\theta^{\top} (t_n - t_{n-1})} + \Omega_{nn} V \\
    \Rightarrow I &= -V^{-1} e^{-\theta(t_n - t_{n-1})} V \left[  V - V e^{\theta^{\top}(t_n - t_{n-1})} V^{-1} e^{-\theta (t_n - t_{n-1})} V \right]^{-1} V e^{-\theta^{\top} (t_n - t_{n-1})} + \Omega_{nn} V \\
    \Rightarrow \Omega_{nn} &= V^{-1} + V^{-1} e^{-\theta(t_n - t_{n-1})} V \left[ V - V e^{-\theta^{\top} (t_n - t_{n-1})} V^{-1} e^{-\theta (t_n - t_{n-1})} V \right]^{-1} V e^{-\theta^{\top} (t_n - t_{n-1})} V^{-1}
\end{aligned}
\end{equation*}

Thus, the precision matrix $\Omega$ is block tri-diagonal with the following entries (indexed by j) for $1 < j < n$:
\begin{equation*}
\begin{aligned}
    & V := vec^{-1} \big\{ (\theta \oplus \theta)^{-1} vec\{\sigma \sigma^{\top}\} \big\} \\
    & \Omega_{11} = \big[ V - V  e^{-\theta^{\top} (t_2 - t_1)} V^{-1} e^{-\theta (t_2 - t_1)} V \big]^{-1} \\
    & \Omega_{j,j+1} = \Omega_{j+1,j}^{\top} = -\big[V - V e^{-\theta^{\top} (t_{j+1} - t_j)} V^{-1} e^{-\theta (t_{j+1} - t_j)}  V \big]^{-1}  V e^{-\theta^{\top} (t_{j+1} - t_j)} V^{-1} \\
    & \Omega_{jj} = V^{-1} + V^{-1} e^{-\theta (t_j - t_{j-1})} V \big[V - V e^{-\theta^{\top} (t_j - t_{j-1})} V^{-1} e^{-\theta (t_j - t_{j-1})} V \big]^{-1} V e^{-\theta^{\top} (t_j - t_{j-1})} V^{-1} \\ & \hspace{0.9cm} + \big[V - V e^{-\theta^{\top} (t_{j+1} - t_j)} V^{-1} e^{-\theta (t_{j+1} - t_j)} V \big]^{-1} V e^{-\theta^{\top} (t_{j+1} - t_j)} V^{-1} e^{-\theta (t_{j+1} - t_j)} \\
    & \Omega_{nn} = V^{-1} + V^{-1} e^{-\theta (t_n - t_{n-1})} V \big[V - V e^{-\theta^{\top} (t_n - t_{n-1})} V^{-1} e^{-\theta (t_n - t_{n-1})} V \big]^{-1} V e^{-\theta^{\top} (t_n - t_{n-1})} V^{-1}
\end{aligned}
\end{equation*}

\section{Identifiability constraint: re-scaling the OU process}\label{ss:rescaling_OU}

Let $(\theta^*, \sigma^*)$ be a pair of OU process parameters satisfying the identifiability constraint that the stationary variance of the OU process is equal to 1; that is, $diag\{\Psi(\theta^*, \sigma^*)\} = 1$, where $\Psi$ is the covariance matrix of the OU process. We show that we can always find a pair of $(\theta^*, \sigma^*)$ that defines a valid mean-reverting OU process with stationary variance of 1 that has the same correlation structure as the original unconstrained OU process defined by ($\theta$, $\sigma$). As an example, consider the stochastic differential equation definition of the bivariate OU process. For an arbitrary mean-reverting OU process, $\eta(t)$, 
\begin{equation*}
    d\begin{bmatrix}
     \eta_1(t) \\ \eta_2(t)
    \end{bmatrix} =
    - \begin{bmatrix}
    \theta_{11} & \theta_{12} \\ \theta_{21} & \theta_{22}
    \end{bmatrix}
    \begin{bmatrix}
    \eta_1(t) \\ \eta_2(t)
    \end{bmatrix} dt + 
    \begin{bmatrix}
    \sigma_{11} & 0 \\ 0 & \sigma_{22}
    \end{bmatrix}
    d \begin{bmatrix}
    W_1(t) \\ W_2(t)
    \end{bmatrix}
\end{equation*}

We could equivalently define this OU process $\eta(t)$ as
\begin{equation*}
\begin{aligned}
    d\begin{bmatrix}
     \eta_1(t) \\ \eta_2(t)
    \end{bmatrix} &=
    - \begin{bmatrix}
    \theta_{11} & \theta_{12} \\ \theta_{21} & \theta_{22}
    \end{bmatrix}
    \begin{bmatrix}
        c_1 & 0 \\ 0 & c_2
    \end{bmatrix}
    \begin{bmatrix}
        1/c_1 & 0 \\ 0 & 1/c_2
    \end{bmatrix}
    \begin{bmatrix}
    \eta_1(t) \\ \eta_2(t)
    \end{bmatrix} dt + 
    \begin{bmatrix}
    \sigma_{11} & 0 \\ 0 & \sigma_{22}
    \end{bmatrix}
    d \begin{bmatrix}
    W_1(t) \\ W_2(t)
    \end{bmatrix} \\
    &= - \begin{bmatrix}
    c_1\theta_{11} & c_2\theta_{12} \\ c_1\theta_{21} & c_2\theta_{22}
    \end{bmatrix}
    \begin{bmatrix}
    \frac{1}{c_1}\eta_1(t) \\ \frac{1}{c_2}\eta_2(t)
    \end{bmatrix} dt + 
    \begin{bmatrix}
    \sigma_{11} & 0 \\ 0 & \sigma_{22}
    \end{bmatrix}
    d \begin{bmatrix}
    W_1(t) \\ W_2(t)
    \end{bmatrix}
\end{aligned}
\end{equation*}

Let $\eta^*(t)$ be a scaled version of $\eta$ where
$$\begin{bmatrix}
\eta^*_1(t) \\ \eta^*_2(t) \end{bmatrix} = \begin{bmatrix}
\frac{1}{c_1}\eta_1(t) \\ \frac{1}{c_2}\eta_2(t)
\end{bmatrix}$$  and $$\begin{bmatrix} \theta^*_{11} & \theta^*_{12} \\ \theta^*_{21} & \theta^*_{22} \end{bmatrix} = \begin{bmatrix} c_1\theta_{11} & c_2\theta_{12} \\ c_1\theta_{21} & c_2\theta_{22} \end{bmatrix}$$

and assume that $\eta^*(t)$ has a stationary variance equal to 1. Then,
\begin{equation*}
\begin{aligned}
    d \eta^*(t) &= - \begin{bmatrix} \theta^*_{11} & \theta^*_{21} \\ \theta^*_{12} & \theta^*_{22} \end{bmatrix} \begin{bmatrix} \eta^*_1(t) \\ \eta^*_2(t) \end{bmatrix} dt + \begin{bmatrix} \sigma^*_{11} & 0 \\ 0 & \sigma^*_{22} \end{bmatrix} d \begin{bmatrix} W_1(t) \\ W_2(t) \end{bmatrix} \\
    &= - \begin{bmatrix} \theta^*_{11} & \theta^*_{21} \\ \theta^*_{12} & \theta^*_{22} \end{bmatrix} \begin{bmatrix} c_1 & 0 \\ 0 & c_2 \end{bmatrix} \begin{bmatrix} \eta_1(t) \\ \eta_2(t) \end{bmatrix} dt + \begin{bmatrix} \sigma^*_{11} & 0 \\ 0 & \sigma^*_{22} \end{bmatrix} d \begin{bmatrix} W_1(t) \\ W_2(t) \end{bmatrix} \\
    &= - \begin{bmatrix} c_1\theta^*_{11} & c_2\theta^*_{21} \\ c_1\theta^*_{12} & c_2\theta^*_{22} \end{bmatrix} \begin{bmatrix} \eta_1(t) \\ \eta_2(t) \end{bmatrix} dt + \begin{bmatrix} \sigma^*_{11} & 0 \\ 0 & \sigma^*_{22} \end{bmatrix} d \begin{bmatrix} W_1(t) \\ W_2(t) \end{bmatrix}
\end{aligned}
\end{equation*}

Looking back at the original OU process $\eta(t)$,
\begin{equation*}
\begin{aligned}
    d\eta(t) &= d\begin{bmatrix} \frac{1}{c_1} & 0 \\ 0 & \frac{1}{c_2} \end{bmatrix} \eta^*(t) \\
    &= -\begin{bmatrix} \frac{1}{c_1} & 0 \\ 0 & \frac{1}{c_2} \end{bmatrix} \begin{bmatrix}  c_1 \theta^*_{11} & c_2\theta^*_{12} \\ c_1 \theta^*_{21} & c_2 \theta^*_{22} \end{bmatrix} \eta(t) dt + \begin{bmatrix} \frac{1}{c_1} \sigma^*_{11} & 0 \\ 0 & \frac{1}{c_2}\sigma^*_{22} \end{bmatrix} dW(t) \\
    &= - \begin{bmatrix} \frac{c_1}{c_1} \theta^*_{11} & \frac{c_2}{c_1} \theta^*_{12} \\ \frac{c_1}{c_2} \theta^*_{21} & \frac{c_2}{c_2} \theta^*_{22} \end{bmatrix} \eta(t) dt + \begin{bmatrix} \frac{1}{c_1} \sigma^*_{11} & 0 \\ 0 & \frac{1}{c_2} \sigma^*_{22} \end{bmatrix} dW(t)
\end{aligned}
\end{equation*}

Finally, we see that the parameters for $\eta(t)$ can easily be re-scaled to satisfy our identifiability assumption:
\begin{equation*}
    \begin{bmatrix} \theta_{11} & \frac{c_1}{c_2} \theta_{12} \\ \frac{c_2}{c_1} \theta_{21} & \theta_{22} \end{bmatrix} = \begin{bmatrix} \theta^*_{11} & \theta^*_{12} \\ \theta^*_{21} & \theta^*_{22} \end{bmatrix}
\end{equation*}

and
\begin{equation*}
    \begin{bmatrix} c_1 \sigma_{11} & 0 \\ 0 & c_2 \sigma_{22} \end{bmatrix} = \begin{bmatrix} \sigma^*_{11} & 0 \\ 0 & \sigma^*_{22} \end{bmatrix}
\end{equation*}

Thus, we have shown that for a mean-reverting bivariate OU process defined by $\theta$ and $\sigma$ with covariance matrix $\Psi(\theta, \sigma)$ and correlation matrix $\Psi^*(\theta, \sigma)$, we can re-scale this OU process to have stationary variance equal to 1 by scaling $\theta_{12}, \theta_{21}$ and $\sigma_{11}, \sigma_{22}$ by a pair of positive scalar constants, $(c_1, c_2)$. This proof can easily be extended to higher dimensional OU processes.

\section{Derivation of the analytic gradients for the measurement submodel}\label{fa_grads}

We have previously defined the log-likelihood for a single subject $i$ as
\begin{equation}
\begin{aligned}
    \ell_i = -\frac{1}{2}log|\Sigma_i^{*}| + Y_i^{\top} \Sigma_i^{*-1} Y_i
\end{aligned}
\end{equation}

where we ignore the constant terms and
\begin{equation}
\begin{aligned}
    \Sigma_i^* = (I_{n_i} \otimes \Lambda) \Psi_i (I_{n_i} \otimes \Lambda)^{\top} +  J_{n_i} \otimes \Sigma_{u} +  I_{n_i} \otimes \Sigma_{\epsilon}
\end{aligned}
\end{equation}

\paragraph{Gradient w.r.t. the loadings:}

We first take the derivative of $\ell_i$ with respect to the elements of the loadings matrix $\Lambda$, $\lambda_k$, $k = 1, ...., p \times K$. The first element of the loadings matrix is parameterized on the log scale in order to restrict this element to positive values for identifiability purposes and so the gradient of this element looks slightly different.  For $k > 1$, we have
\begin{equation}
\begin{aligned}
\label{eq:grad_lambda}
    \frac{\partial \ell_i}{\partial \lambda_k} = -\frac{1}{2} \Bigg[ tr\Big\{ \Sigma_i^{*-1} \frac{\partial \Sigma_i^*}{\partial \lambda_k} \Big\} - Y_i^{\top} \Sigma_i^{*-1} \frac{\partial \Sigma_i^*}{\partial \lambda_k} \Sigma_i^{*-1} Y_i \Bigg]
\end{aligned}
\end{equation}

where
\begin{equation}
\begin{aligned}
    \frac{\partial \Sigma_i^*}{\partial \lambda_k} = (I_{n_i} \otimes \Lambda) \Psi_i (I_{n_i} \otimes J^{k^{\top}}) + (I_{n_i} \otimes J^k) \Psi_i (I_{n_i} \otimes \Lambda^{\top})
\end{aligned}
\end{equation}

We use $J^{k}$ as an indicator matrix that has the same dimension as $\Lambda$ but contains all zeros except for a single 1 indicating the location of element $\lambda_k$.  For $k = 1$, we apply the chain rule and have
\begin{equation}
\begin{aligned}
    \frac{\partial \ell_i}{\partial log(\lambda_k)} = \frac{\partial \ell_i}{\partial \lambda_k} \Bigg[ \frac{\partial log(\lambda_k)}{\partial \lambda_k} \Bigg]^{-1} = \frac{\partial \ell_i}{\partial \lambda_k} \lambda_k
\end{aligned}
\end{equation}

\paragraph{Gradient w.r.t. the random effects:}

Next, we take the gradient of $\ell_i$ with respect to the elements of $R_u$ where $R_u$ comes from the Cholesky decomposition of the random effects covariance matrix, $\Sigma_u = R_u^{\top}R_u$.  For $p,q = 1, ..., K$ and $p \neq q$,
\begin{equation}
\begin{aligned}
    \frac{\partial \Sigma_i^*}{\partial r_{pq}} = J_{n_i} \otimes (J^{k^{\top}}R_u + R_u^{\top} J^k)
\end{aligned}
\end{equation}

\begin{equation}
\begin{aligned}
\label{eq:grad_sigma_u}
    \frac{\partial \ell_i}{\partial r_{pq}} = -\frac{1}{2} \Bigg[ tr\Bigg\{ \Sigma_i^{*-1} \frac{\partial \Sigma_i^*}{\partial r_{pq}} \Bigg\} + Y_i^{\top} \Sigma_i^{*-1} \frac{\partial \Sigma_i^*}{\partial r_{pq}} \Sigma_i^{*-1} Y_i \Bigg]
\end{aligned}
\end{equation}

where again $J^k$ is an indicator matrix of the same dimensions as $\Sigma_u$.

For $p,q = 1, ..., K$ and $p = q$,
\begin{equation}
\begin{aligned}
    \frac{\partial \ell_i}{\partial log(r_{u_{pp}})} = \frac{\partial \ell_i}{\partial r_{u_{pp}}} \Bigg[ \frac{\partial log(r_{u_{pp}})}{\partial r_{u_{pp}}} \Bigg]^{-1} = \frac{\partial \ell_i}{\partial r_{u_{pp}}} r_{u_{pp}}
\end{aligned}
\end{equation}

Note that if we assume only random intercepts (i.e., a diagonal covariance matrix) then we can avoid the Cholesky decomposition by estimating $\sigma_u$ on the log scale. In this case, the gradient simplifies to the form given below for the measurement error.

\paragraph{Gradient w.r.t. the measurement error:}
\label{sss:grad_sigma_e}

Finally, we take the gradient of $\ell_i$ with respect to the elements of the measurement error covariance matrix, $\Sigma_{\epsilon}$. For $k = 1, ..., K$, we have
\begin{equation}
\begin{aligned}
    \frac{\partial \Sigma_i^*}{\partial \sigma_{\epsilon_{k}}} = I_{n_i} \otimes 2 \sigma_{\epsilon_k} J^k
\end{aligned}
\end{equation}

\begin{equation}
\begin{aligned}
\label{eq:grad_sigma_e}
    \frac{\partial \ell_i}{\partial \sigma_{\epsilon_{k}}} &= -\frac{1}{2} \Bigg[ tr \Bigg\{ \Sigma_i^{*-1} \frac{\partial \Sigma_i^*}{\partial \sigma_{\epsilon_{k}}} \Bigg\} - Y_i^{\top} \Sigma_i^{*-1} \frac{\partial \Sigma_i^*}{\partial \sigma_{\epsilon_{k}}} \Sigma_i^{*-1} Y_i \Bigg]
\end{aligned}
\end{equation}

\begin{equation}
\begin{aligned}
    \frac{\partial \ell_i}{\partial log(\sigma_{\epsilon_k})} = \frac{\partial \ell_i}{\partial \sigma_{\epsilon_k}} \Bigg[ \frac{log(\sigma_{\epsilon_k})}{\partial \sigma_{\epsilon_k}} \Bigg]^{-1} = \frac{\partial \ell_i}{\partial \sigma_{\epsilon_k}} \sigma_{\epsilon_k}
\end{aligned}
\end{equation}

where $J^k$ is an indicator matrix of the same dimensions as $\Sigma_{\epsilon}$.






\section{Parameterization of the log-likelihood for standard error estimation}\label{sup:loglik_se_param}

To make our OUF model identifiable, we impose a constraint on the scale of the OU process by forcing the stationary variance equal to 1 via a set of $p$ positive scalar constants.  These constants are functions of OU parameters $\theta$ and $\sigma$.

When the log-likelihood is allowed to vary as a function all parameters, rather than just a single block of parameters as in our block coordinate descent algorithm, our model is no longer identifiable.  To estimate standard errors, we take advantage of the fact that under the identifiability constraint, $\sigma$ can be written as a function of $\theta$, as shown here:

Recall that the stationary variance of the OU process is $V := vec^{-1}\Big\{ (\theta \oplus \theta)^{-1} vec\{\sigma \sigma^{\top} \} \Big\}$.  Assuming a bivariate OU process, under the identifiability constraint, $V$ takes the form $\begin{bmatrix} 1 & \rho \\ \rho & 1 \end{bmatrix}$ where the off-diagonal element $\rho$ is the correlation.  Then,
\begin{equation*}
\begin{aligned}
    \begin{bmatrix}
    1 & \rho \\ \rho & 1
    \end{bmatrix} = vec^{-1} \Big\{ (\theta \oplus \theta)^{-1} vec\{ \sigma \sigma^{-1} \} \Big\} \Longrightarrow \begin{bmatrix}
    1 \\ \rho \\ \rho \\ 1
    \end{bmatrix} = (\theta \oplus \theta)^{-1} \begin{bmatrix}
    \sigma_{11}^2 \\ 0 \\ 0 \\ \sigma_{22}^2
    \end{bmatrix}.
\end{aligned}
\end{equation*}

Letting $$(\theta \oplus \theta)^{-1} = \begin{bmatrix} x_{11} & x_{12} & x_{13} & x_{14} \\ x_{21} & x_{22} & x_{23} & x_{24} \\ x_{31} & x_{32} & x_{33} & x_{34} \\ x_{41} & x_{42} & x_{43} & x_{44} \end{bmatrix},$$ where each element $x_{ij}$ is some function of the elements of $\theta$, we can solve for $\sigma$ in the ($\theta$, $\sigma$) pair that satisfies the identifiability constraint via
\begin{equation*}
\begin{aligned}
    1 &= x_{11} \sigma^2_{11} + x_{14} \sigma^2_{22} \\
    1 &= x_{41} \sigma^2_{11} + x_{44} \sigma^2_{22}
\end{aligned}
\end{equation*}

By constraining $\sigma$ to be a function of $\theta$, we take an alternative approach to identification and no longer require use of the scaling constants here.

\section{Calculating AIC and BIC}\label{ss:aic_formula}

To compare models fit with different numbers of latent factors, we use AIC and BIC.  The formulas for AIC and BIC are given below.  Note that these formulas account for the identifiability constraints described above in Section \ref{ss:rescaling_OU} and in Lemma 4 of the main text.  Letting $\hat{L}$ denote the maximized value of the likelihood of the OUF model used to calculate point estimates; $q$ be the total number of non-zero parameters in $\Lambda, \Sigma_u, \Sigma_{\epsilon}, \theta_{OU}$ and $\sigma_{OU}$; and $p$ be the number of latent factors (which corresponds to the number of scaling constants needed to impose the identifiability constraint), AIC is calculated as:

\begin{equation*}
\begin{aligned}
    2 \times (q-p) - 2log\hat{L}
\end{aligned}
\end{equation*}

\noindent BIC is calculated similarly as

\begin{equation*}
\begin{aligned}
    2 \times log(N) \times (q-p) - 2log\hat{L}
\end{aligned}
\end{equation*}
where $N$ is the total number of independent subjects in the data.

\section{Choice of true OU process in simulation study}\label{ss:true_ou}

In the simulation study described in the main text (Section 4.1-4.2), we generate data in three different settings in which the true OU process has varying degrees of auto-correlation.  We present the true OU process parameters here:

\paragraph{Setting 1:}

\begin{align*}
    \theta = \begin{bmatrix}
        1 & 0.6 \\ 4 & 5
    \end{bmatrix} \text{  and  }
    \sigma = \begin{bmatrix}
        1 & 0 \\ 0 & 2
    \end{bmatrix}
\end{align*}

\paragraph{Setting 2:}

\begin{align*}
    \theta = \begin{bmatrix}
        1.0 & 0.4 \\ 1.8 & 3.0
    \end{bmatrix} \text{  and  }
    \sigma = \begin{bmatrix}
        1.25 & 0 \\ 0 & 2.00
    \end{bmatrix}
\end{align*}

\paragraph{Setting 3:}

\begin{align*}
    \theta = \begin{bmatrix}
         1 & 0.5 \\ 2 & 5
    \end{bmatrix} \text{  and  }
    \sigma = \begin{bmatrix}
        2 & 0 \\ 0 & 3
    \end{bmatrix}
\end{align*}

\vspace{1cm}

In the simulation study assessing use of AIC and BIC to select the correct number of latent factors in a model (described in the main text in Section 4.3), the true parameters were set to the values listed below.  The true values used for $\Sigma_u$ and $\Sigma_{\epsilon}$ were the same as in the original simulation study (see Section 4.1 in the main text).

\paragraph{One factor model:}

\begin{align*}
    \Lambda = \begin{bmatrix}
    1.2 \\ 1.8\\ -0.4\\ 2
        \end{bmatrix}, \ \ 
    \theta = 0.8, \ \  \sigma = 1
\end{align*}

\paragraph{Two factor model with low signal:}

\begin{align*}
    \Lambda = \begin{bmatrix}
    1.2 & 0 \\ 1.8 & 0 \\ 0 & -0.4 \\ 0 & 2
        \end{bmatrix}, \ \ 
    \theta = \begin{bmatrix}
        2 & 0.5 \\ 0.4 & 4
    \end{bmatrix}, \ \ 
    \sigma = \begin{bmatrix}
        2 & 0 \\ 0 & 1
    \end{bmatrix}
\end{align*}

\paragraph{Two factor model with high signal:}

\begin{align*}
    \Lambda = \begin{bmatrix}
    1.2 & 0 \\ 1.8 & 0 \\ 0 & -0.4 \\ 0 & 2
        \end{bmatrix}, \ \ 
    \theta = \begin{bmatrix}
        1 & 1.5 \\ 2 & 5
    \end{bmatrix}, \ \ 
    \sigma = \begin{bmatrix}
        2 & 0 \\ 0 & 3
    \end{bmatrix}
\end{align*}

\paragraph{Three factor model with low signal:}

\begin{align*}
    \Lambda = \begin{bmatrix}
    1.2 & 0 & 0 \\ 1.8 & 0 & 0 \\ 0 & -0.4 & 0 \\ 0 & 0 & 2
        \end{bmatrix}, \ \ 
    \theta = \begin{bmatrix}
        2 & 0.2 & 0.4 \\ 0.8 & 1.1 & 0.5 \\ 0.7 & 0.5 & 1.2
    \end{bmatrix}, \ \ 
    \sigma = \begin{bmatrix}
        1.2 & 0 & 0 \\ 0 & 0.8 & 0 \\ 0 & 0 & 0.4 
    \end{bmatrix}
\end{align*}

\paragraph{Three factor model with high signal:}

\begin{align*}
    \Lambda = \begin{bmatrix}
    1.2 & 0 & 0 \\ 1.8 & 0 & 0 \\ 0 & -0.4 & 0 \\ 0 & 0 & 2
        \end{bmatrix}, \ \ 
    \theta = \begin{bmatrix}
        1 & 0.4 & 0.6 \\ 1.8 & 3 & 0.9 \\ 0.9 & 1 & 1.2
    \end{bmatrix}, \ \ 
    \sigma = \begin{bmatrix}
        1.2 & 0 & 0 \\ 0 & 0.8 & 0 \\ 0 & 0 & 0.4 
    \end{bmatrix}
\end{align*}

\section{Discussion of numerical issues in simulation results}\label{supp:sim_issues}

\paragraph{Simulation study: settings 1-3}

The estimation algorithm failed to converge due to numerical issues when applied to a few of the simulated datasets generated in the simulation study described in Section 4.1-4.2. The failures were caused by a singular V matrix at the start of the first block update of the structural submodel parameters.  Slightly altering the values at which the OU process parameters were initialized resolved this issue.  Point estimates were ultimately calculate for all 1000 simulated datasets in each setting.  In Setting 3, an invalid variance for the measurement submodel parameter $\sigma_{\epsilon_4}$ was estimated from one dataset. In this instance, the variance estimated for this parameter was negative.  We attribute this issue to the numerical approximation used to calculate the Hessian when applied to these this dataset of size $N = 200$.  We anticipate that a larger dataset would improve the approximation of the numerical Hessian but chose to simulate a dataset of this size in order to assess model performance in a realistic setting similar to that encountered in the motivating data application.  In practical application, if a negative variance were to be estimated, it could be rounded to 0. In the results presented in the main text, we ignore the variance estimate for this one $\sigma_{\epsilon_4}$.

\paragraph{Simulation study: model selection}

In our simulation studies, we aimed to assess simulated datasets with sample sizes similar to that of our motivating dataset.  For datasets of fixed size ($N = 200$ subjects), we found that convergence speeds decrease and estimation becomes more difficult as the number of factors in the model increases.  We found that point estimates of the diagonal elements of $\theta_{OU}$ hit the lower bound of $1\times10^{-4}$ less than 1\% of the time.  To improve convergence, we slightly altered the set of default parameter values considered during the initialization steps of the block-wise estimation algorithm for a subset of datasets.  However, when assessing AIC and BIC as model selection criteria (see Section 4.3), we very occasionally encountered numerical issues and so failed to calculate parameter estimates for a subset of models applied to the simulated datasets.  The results reported in the main paper correspond to a comparison of AIC and BIC across datasets for which the algorithm used to fit all three models (the one-factor, two-factor, and three-factor models) either converged or reached the maximum number of iterations prior to convergence.  We assessed whether or not including results in which the maximum number of iterations was reached prior to convergence impacted our model selection results and found no substantial changes.  Supplementary Table \ref{tab:GoF_results_convg_only} shows the equivalent version of Table 1 presented in the main text if only results from datasets that had converged were shown.

\begin{table}
    \centering
    \includegraphics[width=\linewidth]{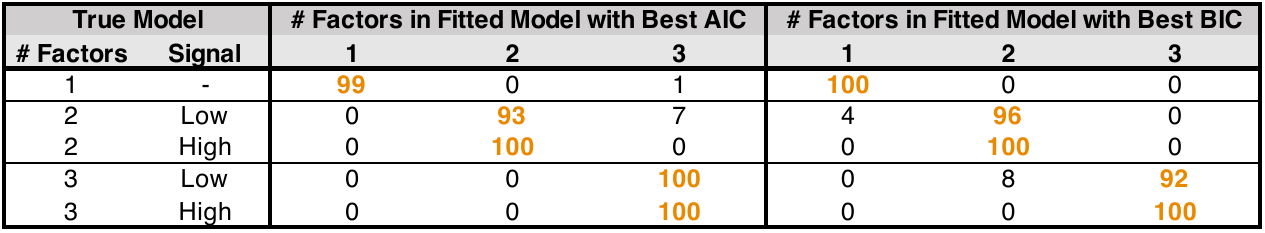}
    \caption{For datasets generated under each true model, we summarize the percent of times that the model-selection metric chose the fitted model with the indicated number of factors.  The settings in which the fitted model has the same number of factors as the true data-generating model are emphasized with bold orange text. These results are presented for datasets on which the algorithm converged prior to reaching the maximum number of iterations (200) for all three models.}\label{tab:GoF_results_convg_only} 
\end{table}

In Supplementary Table \ref{tab:GoF_results_convg_summary}, we summarize the number (out of 100) of datasets (in each setting) for which the algorithm converged (using $\delta = 1\times10^{-6})$ or reached the maximum number of iterations prior to convergence.  When this total number does not add up to 100, the remaining datasets correspond to situations in which the algorithm failed due to numerical issues (e.g., current OU parameter estimates corresponded to a singular stationary covariance matrix).

\begin{table}
    \centering
    \includegraphics[width=\linewidth]{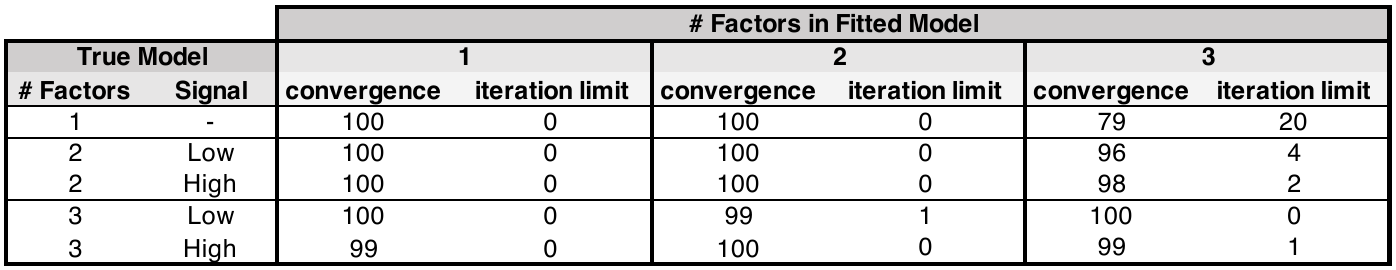}
    \caption{For datasets generated under each true model, we summarize the number of datasets (out of 100) on which the algorithm converged or reached the maximum number of block-wise iterations prior to convergence (when $\delta = 1\times10^{-6}$).  For totals that do not sum to 100, the remaining cases correspond to instances in which the algorithm failed due to numerical issues prior to converging or reaching the maximum number of block-wise iterations (200).}\label{tab:GoF_results_convg_summary} 
\end{table}

After loosening the convergence criteria across the block-wise iterations, we did not find substantially different results when evaluating AIC and BIC as model selection criteria when compared to results under the original convergence criteria.  For example, if we categorized convergence using $\delta \le 1\times10^{-3}$, rather than only the original $\delta = 1\times10^{-6}$, the algorithm would have converged when fitting almost every model to almost every dataset (see Supplementary Table \ref{tab:GoF_results_convg_summary_1e-3}) but the model selection results would not have changed (see Supplementary Table \ref{tab:GoF_results_convg_only_1e-3}).

\begin{table}
    \centering
    \includegraphics[width=\linewidth]{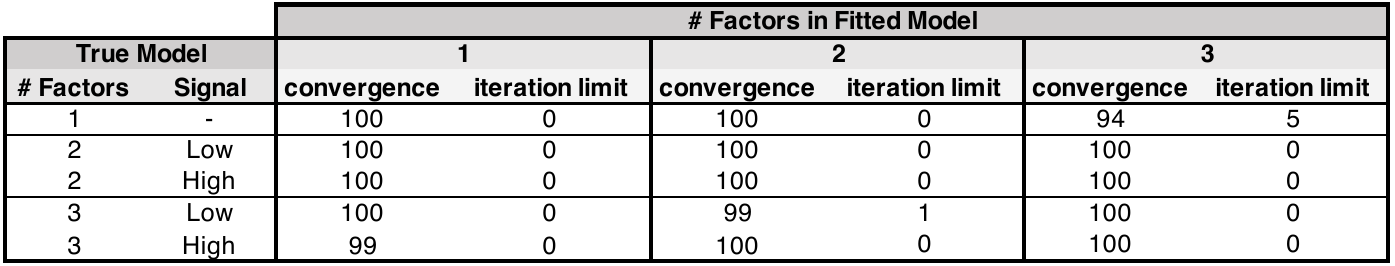}
    \caption{For datasets generated under each true model, we summarize the number of datasets (out of 100) on which the algorithm converged or reached the maximum number of block-wise iterations prior to convergence (when $\delta \le 1\times10^{-3}$).  For totals that do not sum to 100, the remaining cases correspond to instances in which the algorithm failed due to numerical issues prior to converging or reaching the maximum number of block-wise iterations (200).}
    \label{tab:GoF_results_convg_summary_1e-3} 
\end{table}

\begin{table}
    \centering
    \includegraphics[width=\linewidth]{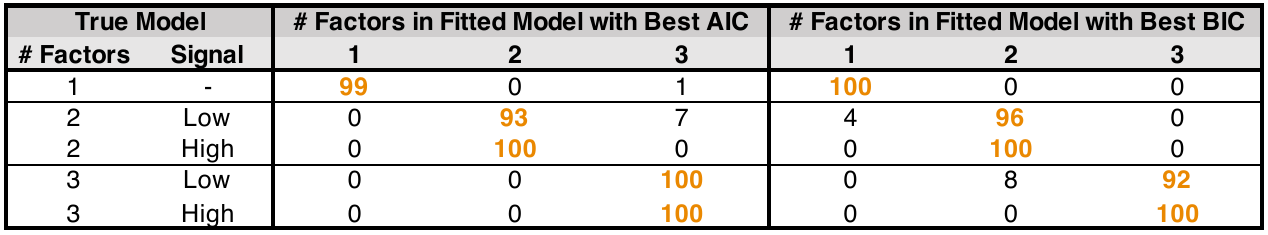}
    \caption{For datasets generated under each true model, we summarize the percent of times that the model-selection metric chose the fitted model with the indicated number of factors.  The settings in which the fitted model has the same number of factors as the true data-generating model are emphasized with bold orange text. These results are presented for datasets on which the algorithm converged (using $\delta \le 1\times10^{-3}$)  prior to reaching the maximum number of iterations (200) for all three models.}\label{tab:GoF_results_convg_only_1e-3} 
\end{table}

We expect that increasing the size of the simulated dataset would increase the rate at which we successfully fit models with more factors.

\renewcommand{\thesection}{B.\arabic{section}}
\setcounter{section}{0}
\section*{Section B}

\section{Application to mHealth emotion data}

\subsection{OUF model with one factor}

In this model, we assume that a single latent factor generates all observed emotions of happy, joyful, enthusiastic, active, calm, determined, grateful, proud, attentive, sad, scared, disgusted, angry, ashamed, guilty, irritable, lonely, and nervous.  We plot the point estimates from this model in Supplementary Figure \ref{fig:point_ests_1factor}.  Using these estimated parameters, we calculate the auto-correlation half-life of this latent factor as approximately 27 days.  This model has a total of 56 free parameters, along with one constraint, which we use when calculating AIC and BIC.

\begin{figure}
    \centering
    \includegraphics[width=\linewidth]{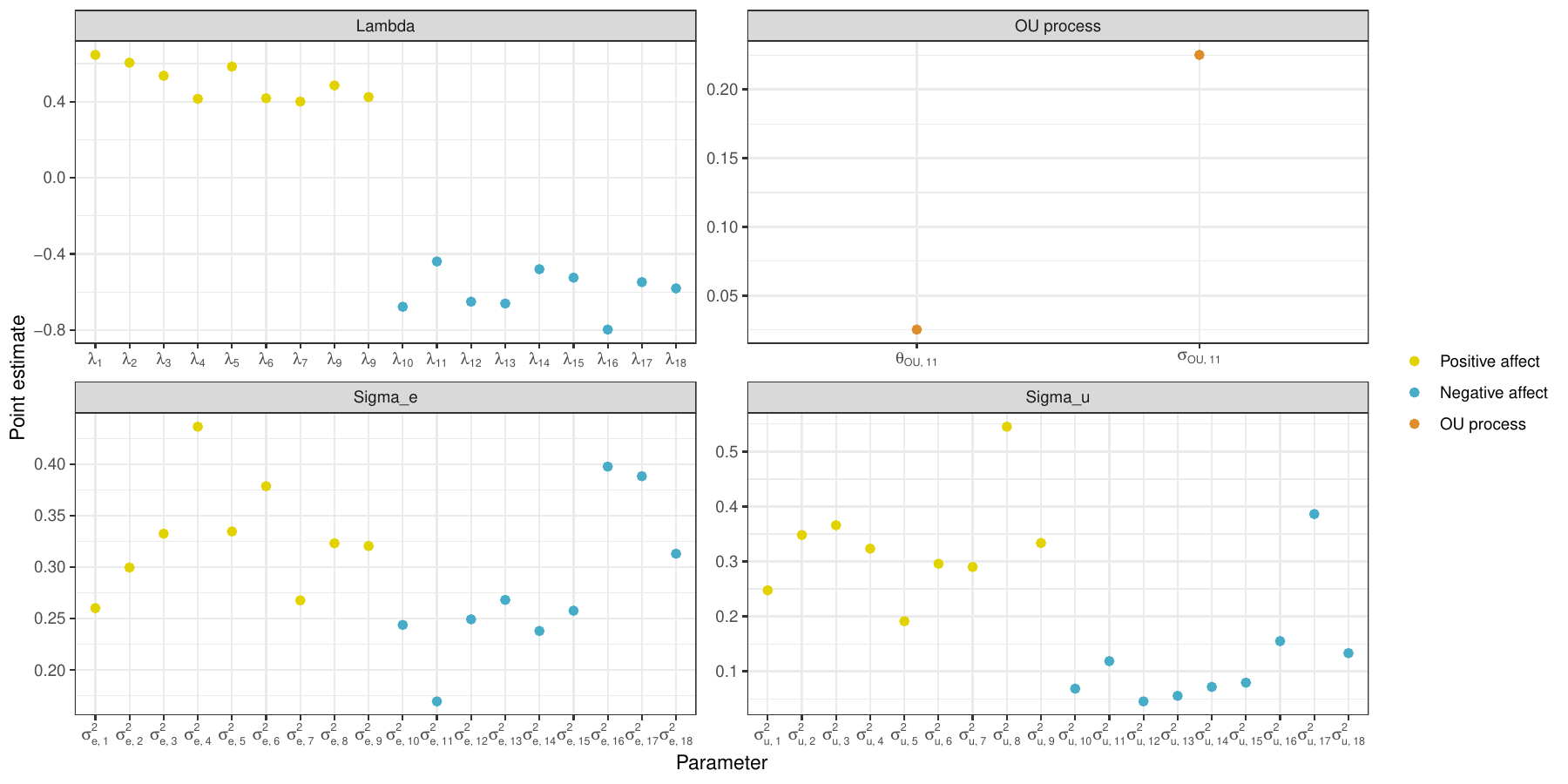}
    \caption{Point estimates for each of the parameter matrices in our one-factor OUF model.  Because we assume structural zeros in the loadings matrix are known, each emotion has only a single loading. Parameter subscripts 1-18 correspond to the emotions as follows: 1 = happy, 2 = joyful, 3 = enthusiastic, 4 = active, 5 = calm, 6 = determined, 7 = grateful, 8 = proud, 9 = attentive, 10 = sad, 11 = scared, 12 = disgusted, 13 = angry, 14 = ashamed, 15 = guilty, 16 = irritable, 17 = lonely, 18 = nervous.}\label{fig:point_ests_1factor}
\end{figure}

\subsection{OUF model with two factors}

In this model, we assume that two latent factors generate the observed emotions.  The latent factors represent positive affect (which underlies happy, joyful, enthusiastic, active, calm, determined, grateful, proud, and attentive) and negative affect (which underlies sad, scared, disgusted, angry, ashamed, guilty, irritable, lonely, and nervous).  Results from this fitted model are available in Section 5 of the main text.  This model has a total of 60 free parameters, along with two constraints, which we use when calculating AIC and BIC.

\subsection{OUF model with three factors}

We assume that three latent emotional states underlie the emotions observed during this study.  The emotions load on to the latent factors as follows:

\begin{enumerate}
    \item enthusiastic, proud, active, calm, determined, attentive, grateful [$\eta_1 = $ high arousal positive affect]
    \item calm, happy, joyful [$\eta_2 = $ no-to-low arousal positive affect]
    \item sad, scared, disgusted, angry, ashamed, guilty, irritable, lonely, nervous [$\eta_3 = $ negative affect]
\end{enumerate}

We use behavioral science literature and theory---namely the circumplex model of emotion---to inform the division of the positive affect emotions into groups representing high arousal positive affect and no-to-low arousal positive affect (see \cite{reisenzein_1994, remington_2000, gilbert_2008, mcmanus_2019}).  Literature supporting the placement of each positive affect emotion is summarized in Table \ref{tab:refs_for_3factor_mod}.  Happy and joyful are also commonly placed midway between high and low arousal in the circumplex model of emotion (see \cite{remington_2000}) and so we chose to assess the fit of the OUF model when these emotion items load onto the latent factor representing no-to-low arousal positive affect.  This model converged after 211 block iterations and we present point estimates in Supplementary Figure \ref{fig:point_ests_3factor}.  This model has a total of 66 free parameters, along with three constraints, which we use when calculating AIC and BIC.

\begin{table}
    \centering
    \includegraphics[width=0.7\linewidth]{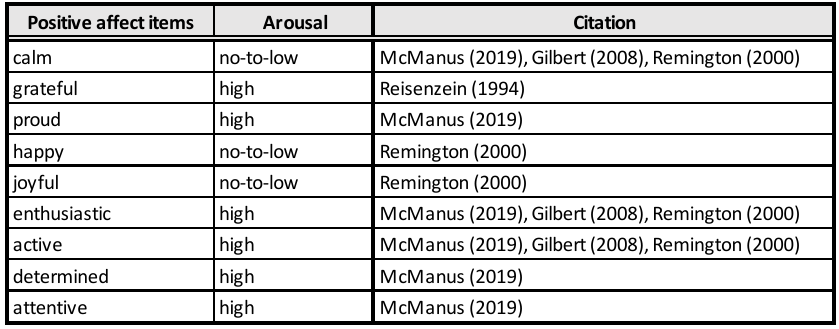}
    \caption{Behavioral science literature supporting the division of the positive emotions into two groups representing no-to-low arousal positive affect and high arousal positive affect.}
    \label{tab:refs_for_3factor_mod} 
\end{table}

\begin{figure}
    \centering
    \includegraphics[width=\linewidth]{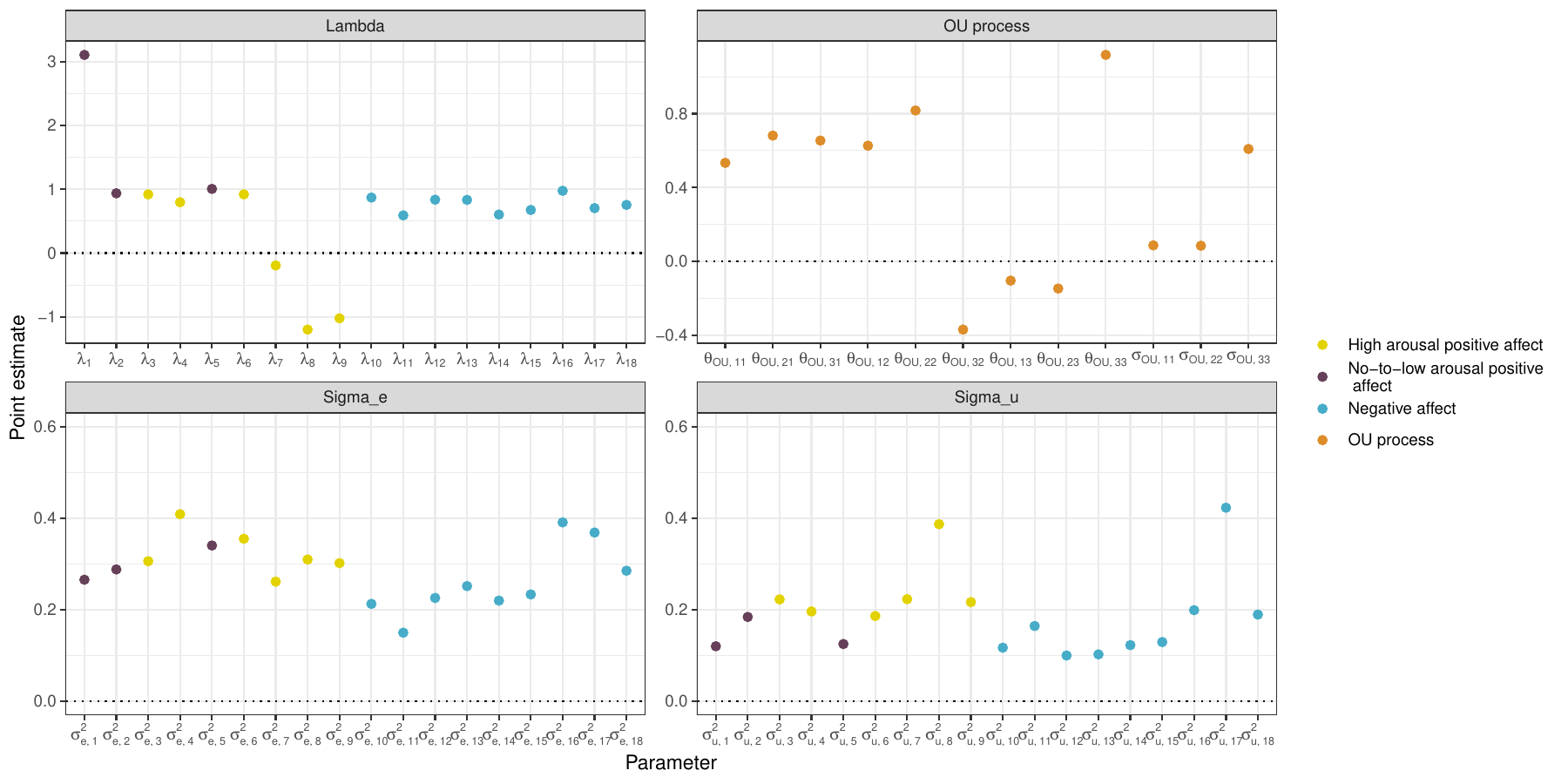}
    \caption{Point estimates for each of the parameter matrices in our three-factor OUF model.  Because we assume structural zeros in the loadings matrix are known, each emotion has only a single loading. Parameter subscripts 1-18 correspond to the emotions as follows: 1 = happy, 2 = joyful, 3 = enthusiastic, 4 = active, 5 = calm, 6 = determined, 7 = grateful, 8 = proud, 9 = attentive, 10 = sad, 11 = scared, 12 = disgusted, 13 = angry, 14 = ashamed, 15 = guilty, 16 = irritable, 17 = lonely, 18 = nervous.}\label{fig:point_ests_3factor}
\end{figure}


\renewcommand{\thesection}{C.\arabic{section}}
\setcounter{section}{0}
\section*{Section C}

\section{Estimation algorithm}\label{sup:est_alg}

\subsection{Parameter initialization}

Due to the complexity of our model, our estimation algorithm is sensitive to the choice of initial estimates. Here we present an approach to estimating reasonable starting values based on simple existing models prior to maximizing the entire likelihood.

\begin{enumerate}
    \item To initialize the \textbf{measurement submodel parameters}, fit a standard cross-sectional factor model to the data collapsed across time (do not include a random intercept but do assume that the positions of the non-zero loadings are known).
    \item Using this fitted factor model, estimate the factor scores (predicted values for $\eta_1$ and $\eta_2$). 
    \item Fit four separate linear mixed effects models---one for each of the observed outcomes, $Y_1, ..., Y_K$---including the factor scores as fixed effects and a random intercept for subject. We do not include a fixed effect intercept in these models. For outcome $k = 1, ..., K$, subject $i = 1, ..., N$, and measurement occasion $j = 1, ..., n_i$, the mixed model is given by $$Y_{kij} = \lambda_{k} \eta_i(t_j) + u_{k0i} + \epsilon_{kij}$$ where $u_{k0i} \sim N(0, \sigma^2_{u_k})$ and $\epsilon_{kij} \sim N(0, \sigma^2_{\epsilon_k}$).
    \item From each of these $K$ mixed models, extract estimates of the coefficient for the fixed effect, the variance for the random intercept, and the residual variance.  Use the coefficients of the fixed effects to initialize the non-zero elements of $\Lambda$ and the variance estimates to initialize the diagonal components of $\Sigma_u$ and $\Sigma_{\epsilon}$.  In some cases, the estimated variances were very small, so a lower limit of 0.1 was set for the initial parameter values to avoid extremely negative estimates after logging. We also set the same lower bound for initial values of the elements in the loadings matrix.
    \item To initialize the \textbf{structural submodel parameters}, we add a term for white noise to the OU process likelihood. This noise term will absorb some of the extra variability in the predicted factor scores and allow for more stable estimation. Let $\Gamma_i$ be white noise, then $\eta_i \sim N(0, \Psi_i + \Gamma_i)$ where $\Gamma_i$ is a diagonal matrix (of the same dimension as OU covariance matrix $\Psi_i$) with constant but unknown diagonal $\gamma$. We then maximize this likelihood and use the estimated OU process parameter values as initial values, restricting the maximum initial values of the diagonals of $\theta_{OU}$ to be less than 7.  This maximum helps deal with instability in the initial estimate of $\theta$.
\end{enumerate}

\subsection{Maximization of the marginal log-likelihood}

To maximize the log-likelihood, we use quasi-Newton optimizers as implemented in the \texttt{stats} package in \texttt{R} \citep{R_stats}. To prevent the parameter estimates from diverging to infinite values, we set the maximum allowed step size to 10.

Using the initial parameter values estimated via the approach described in the previous section, we iteratively update measurement and structural submodel parameter estimates in blocks:

\begin{enumerate}
    \item Initialize estimates: $\Lambda^{(0)}, \Sigma_u^{(0)}, \Sigma_{\epsilon}^{(0)}, \theta^{(0)}, \sigma^{(0)}$.  Measurement submodel parameters are always initialized empirically; for structural submodel parameters, two sets of initial estimates are considered---an empirical set of values estimated as described above and a default set of values that are based on a reasonable guess.  The set of values that corresponds to the higher log-likelihood is used.
    \item Set $r = 1$ and $\delta = 0$. While $r \le 200$ and $\delta = 0$,
    \begin{enumerate}
        \item Update block of \textbf{measurement submodel parameters}: $$\Lambda^{(r)}, \Sigma_u^{(r)}, \Sigma_{\epsilon}^{(r)} = \underset{\Lambda, \Sigma_u, \Sigma_{\epsilon}}{argmax}\big\{ logL(Y | \theta^{(r-1)}, \sigma^{(r-1)}) \big\}.$$ We solve this iteratively using \texttt{nlm} \citep{R_stats} and analytic gradients with convergence criteria set to \texttt{gradtol} $= max(1\times10^{-4}/10^r, 1\times10^{-8})$ and \texttt{steptol} $= max(1\times10^{-4}/10^r, 1\times10^{-8})$.  \texttt{gradtol} is the tolerance for the scaled gradient and \texttt{steptol} is the tolerance for parameter estimates across iterations.  We model the first element of the loadings matrix and the variance parameters on the log scale, since all of these estimates are required to be positive.
        \item Update block of \textbf{structural submodel parameters}: $$\theta^{(r)}, \sigma^{(r)} = \underset{\theta, \sigma}{argmax}\big\{ logL(Y | \Lambda^{(r)}, \Sigma_u^{(r)}, \Sigma_{\epsilon}^{(r)}) \big\}.$$ We solve this iteratively using \texttt{nlminb} and numeric approximations to the gradients.  For estimates of $\theta$, the diagonal elements must be positive and the matrix must have eigenvalues with positive real parts.  The diagonal element of $\sigma$ are estimated on the log scale, since they are required to be positive.
        \item Check for block-wise convergence: Let $\Theta$ be a vector containing all elements of $\Lambda$, $\Sigma_u$, $\Sigma_{\epsilon}$, $\theta$, and $\sigma$. Then, calculate $$\delta = \max \Big\{I\big\{|\Theta^{(r)} - \Theta^{(r-1)}|/\Theta^{(r)} < 10^{-6}\big\},  I\big\{logL(\Theta^{(r)} | Y) - logL(\Theta^{(r-1)} | Y) < 10^{-6}\big\}\Big\}$$ where all operations on $\Theta$ are element-wise.
        \item Rescale OU process parameters so stationary variance is equal to 1 using Lemma 4.
        \item Update $r$: $r = r+1$
    \end{enumerate}
    \item Estimate standard errors using a numerical approximation to the Hessian of the joint negative log-likelihood for $\Lambda^{(r)}, \Sigma_u^{(r)}, \Sigma_{\epsilon}^{(r)}, \theta^{(r)}$ at the current parameter values.  Rather than rescaling the OU parameters so stationary variance is equal to 1 using Lemma 4, we assume that $\sigma$ is a function of $\theta$. See Section A.6 of the supplementary material for further description of this function. The numeric approximation to the Hessian is carried out using the \texttt{optimHess} function in the \texttt{stats} package.
    \item Estimate confidence interval for OU process parameter $\sigma$ based on a parametric bootstrap of $\theta$.
\end{enumerate}

\section{Comparison with Tran et al. (2021b)} 

To illustrate the computational benefits of our proposed block coordinate descent algorithm for estimation relative to the Bayesian approach taken in \cite{tran_2021b}, we apply both methods to simulated datasets.  Because we only consider continuous outcomes in this work, we slightly modify the original model proposed in \cite{tran_2021b} and do not estimate the additional parameters used to account for non-continuous outcomes.  \cite{tran_2021b} also consider two different sets of constraints on the OU process drift matrix (denoted here as $\theta_{OU}$); we use the set of constraints that specify the eigenvalues of $\theta_{OU}$ to have positive real parts.

We use the same simulation set-up as described in the main text (Section 4.1) with the true OU process parameters corresponding to Setting 1 (Section \ref{ss:true_ou}).  We make one modification to the true values of the loadings parameters: we restrict \textit{all} elements of the loadings matrix to be positive.  This restriction means that $\lambda_3 = 0.4$, rather than the original $\lambda_3 = -0.4$.  We make this assumption in order to make identification of parameters more straightforward in this comparison of methods.

We generate 100 replicates of the simulated dataset and fit the OUF model using our proposed estimation algorithm and the algorithm proposed in \cite{tran_2021b}.  \cite{tran_2021b} use a slightly different parameterization of the OU process than we use in this work. 
 In our implementation of the OU process, we restrict the volatility parameter matrix, $\sigma_{OU}$, to be a diagonal matrix.  Although \cite{tran_2021b} do not make this assumption, there is still a one-to-one correspondence between the set of parameters estimated in our work and the set of posterior estimates resulting from their Bayesian method.  As a result of these differences in parameters, we do not report estimates of $\sigma_{OU}$ in the plot below and instead present parameter estimates for $\rho$, which is the stationary correlation between $\eta_1$ and $\eta_2$.  Tran et al. estimate this parameter directly and we can calculate an estimate for it using $\hat{\theta}_{OU}$ and $\hat{\sigma}_{OU}$.

\begin{figure} 
    \centering
    \includegraphics[width=0.5\linewidth]{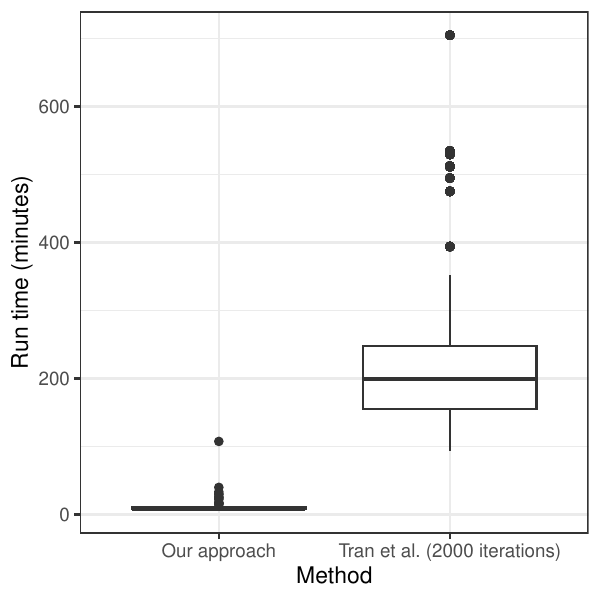}
    \caption{Computation time (in minutes) for the our estimation algorithm and the Bayesian estimation method proposed in \cite{tran_2021b}.  Box plots summarizes the computation time required to fit the OU factor model using both approaches across 100 simulated datasets.  Time required to compute initial parameter estimates is not included in the total above.  For our approach, the total time includes both the time required to carry out the block coordinate descent algorithm plus the time required to estimate standard errors.} \label{fig:comp_time_comparison}
\end{figure}

When applying the Bayesian approach, we use our proposed empirical approach to initializing parameter values, assume 4 chains, and allow the sampler to run for 2,000 iterations.  We discard the first half of these samples as burn-in.  The computation time of both approaches---excluding time required to compute initial parameter estimates---is shown in Figure \ref{fig:comp_time_comparison}.  Computing resources are the same across all replicates; we use 4 cores with a total of 4GB of memory for each replicate (this allows gradients to be evaluated or chains to be sampled in parallel, depending on the method).  We find that our approach, on average, takes approximately 5\% of the time required by the method in Tran et al (2021b).

Point estimates for both estimation approaches are shown in Supplementary Figure \ref{fig:pt_est_comparison}.  We present the posterior means for each parameter across the 100 simulated datasets as estimated using the method from \cite{tran_2021b}; maximum likelihood estimates resulting from our block coordinate descent algorithm are also summarized across the 100 simulated datasets.  Note that the posterior estimates from the Bayesian estimate may be slightly improved by running the sampling algorithm for additional iterations; we limit the MCMC algorithm to 2,000 iterations since our focus is on comparing computation time.

\begin{figure} 
    \centering
    \includegraphics[width=\linewidth]{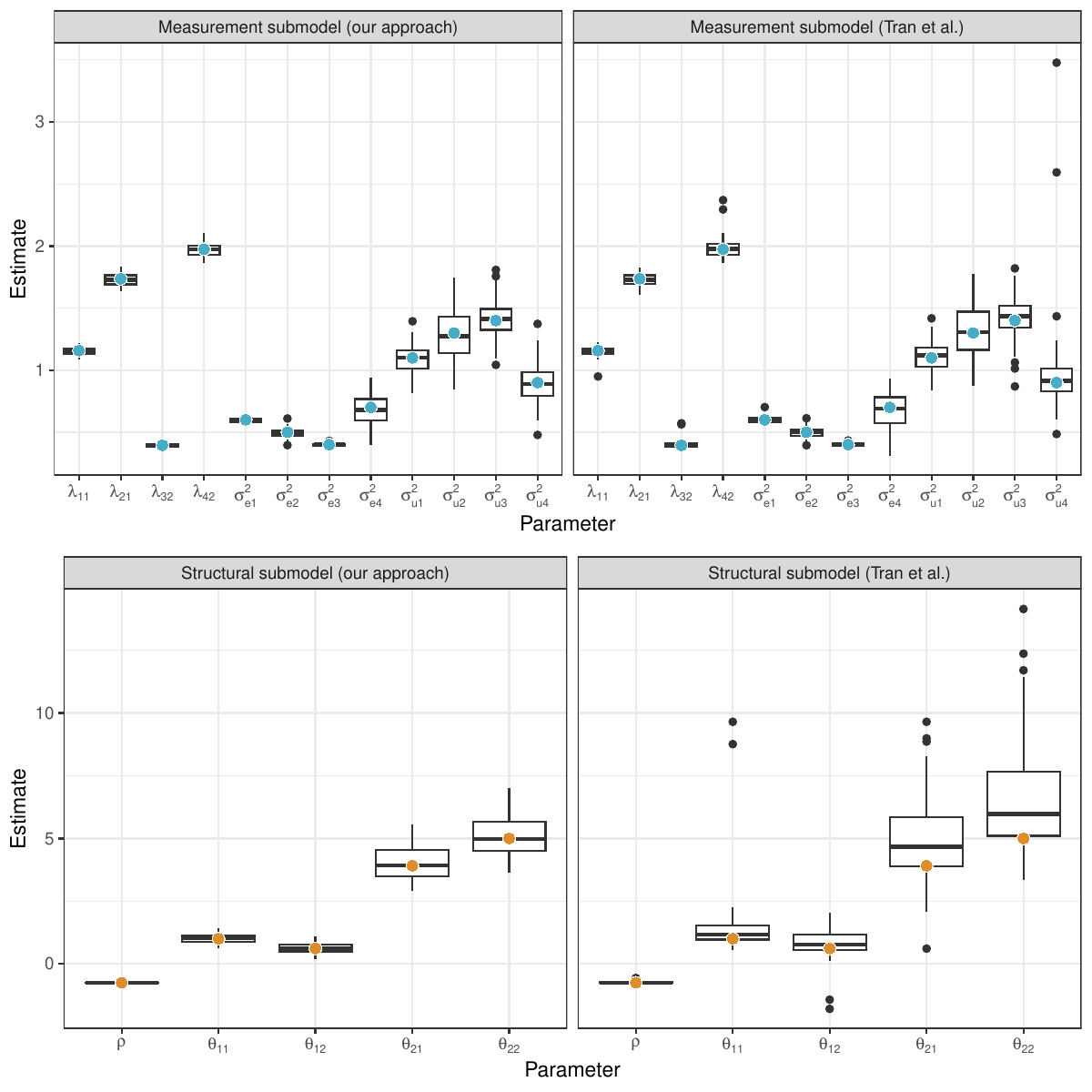}
    \caption{Final parameter estimates from the block coordinate descent algorithm and the Bayesian estimation method proposed in \cite{tran_2021b}.  For the Bayesian method, posterior means are used for point estimates.  Each box plot summarizes point estimates across the 100 simulated datasets.  True parameter values are indicated with colored dots.} \label{fig:pt_est_comparison}
\end{figure}


\newpage 
\bibliographystyle{biom} 
\bibliography{references.bib}